\documentclass[twocolumn]{aastex631}

\usepackage{epsfig}
\usepackage{natbib}
\usepackage{amsmath}
\usepackage{hyperref}
\usepackage{comment}
\usepackage{multirow,tabularx}
\usepackage[flushleft]{threeparttable}

\usepackage{amssymb}
\usepackage{soul}

\usepackage{xcolor}

\begin{document}

\title{The Dark Energy Bedrock All-Sky Supernova Program: Motivation, Design, Implementation, and Preliminary Data Release}

\correspondingauthor{Nora F. Sherman}
\email{norafs@bu.edu}

\author[0000-0001-5399-0114]{Nora F. Sherman}
\affiliation{Institute for Astrophysical Research, Department of Astronomy, Boston University, 725 Commonwealth Avenue, Boston, MA 02215, USA}

\author[0000-0002-5389-7961]{Maria Acevedo}
\affiliation{Department of Physics, Duke University, Durham, NC 27708, USA}

\author[0000-0001-5201-8374]{Dillon Brout}
\affiliation{Institute for Astrophysical Research, Department of Astronomy, Boston University, 725 Commonwealth Avenue, Boston, MA 02215, USA}
\affiliation{Department of Physics, Boston University, 725 Commonwealth Avenue, Boston, MA 02215, USA}

\author[0009-0006-4963-3206]{Bailey Martin}
\affiliation{The Research School of Astronomy and Astrophysics, The Australian National University, Canberra, ACT 2611, Australia}

\author[0000-0002-4934-5849]{Daniel Scolnic}
\affiliation{Department of Physics, Duke University, Durham, NC 27708, USA}

\author[0009-0006-5649-5067]{Dingyuan Cao}
\affiliation{The Research School of Astronomy and Astrophysics, The Australian National University, Canberra, ACT 2611, Australia}

\author[0000-0003-1731-0497]{Christopher Lidman}
\affiliation{The Research School of Astronomy and Astrophysics, The Australian National University, Canberra, ACT 2611, Australia}

\author[0009-0002-6802-8045]{Noor Ali}
\affiliation{Institute of Space Sciences (ICE, CSIC), Campus UAB, Carrer de Can Magrans, s/n, E-08193 Barcelona, Spain}

\author[0000-0003-1997-3649]{Patrick Armstrong}
\affiliation{The Research School of Astronomy and Astrophysics, The Australian National University, Canberra, ACT 2611, Australia}

\author[0000-0002-4449-9152]{K.~Auchettl}
\affiliation{School of Physics, The University of Melbourne, VIC 3010, Australia}
\affiliation{Department of Astronomy and Astrophysics, University of California, Santa Cruz, CA 95064, USA}

\author[0000-0002-4934-5849]{Rebecca C. Chen}
\affiliation{Department of Physics, Duke University, Durham, NC 27708, USA}

\author[0000-0001-8251-933X]{Alex Drlica-Wagner}
\affiliation{Fermi National Accelerator Laboratory, P. O. Box 500, Batavia, IL 60510, USA}
\affiliation{Kavli Institute for Cosmological Physics, University of Chicago, Chicago, IL 60637, USA}
\affiliation{Department of Astronomy and Astrophysics, University of Chicago, Chicago, IL 60637, USA}
\affiliation{NSF-Simons AI Institute for the Sky (SkAI),172 E. Chestnut St., Chicago, IL 60611, USA}

\author[0000-0001-6957-1627]{Peter S. Ferguson}
\affiliation{DiRAC Institute, Department of Astronomy, University of Washington, 3910 15th Ave NE, Seattle, WA, 98195, USA}

\author[0000-0001-6718-2978]{Kenneth Herner}
\affiliation{Fermi National Accelerator Laboratory, P. O. Box 500, Batavia, IL 60510, USA}

\author[0000-0001-6022-0484]{Gautham~Narayan}
\affiliation{University of Illinois, Urbana-Champaign, Dept. of Astronomy
1002  W. Green St., Rm. 127, Urbana, IL 61801}
\affiliation{Center for AstroPhysical Surveys, NCSA
1205 W Clark St, Urbana, IL 61801}

\author[0000-0001-8596-4746]{Erik R. Peterson}
\affiliation{Department of Physics, Duke University, Durham, NC 27708, USA}

\author[0000-0003-0824-669X]{Liana Rauf} \affiliation{The Research School of Astronomy and Astrophysics, The Australian National University, Canberra, ACT 2611, Australia}

\author[0000-0002-4410-5387]{Armin Rest}
\affiliation{Space Telescope Science Institute, 3700 San Martin Drive, Baltimore, MD 21218, USA}
\affiliation{Physics and Astronomy Department, Johns Hopkins University, Baltimore, MD 21218, USA}

\author[0000-0002-6124-1196]{Adam G.~Riess}
\affiliation{Space Telescope Science Institute, Baltimore, MD 21218, USA}
\affiliation{Department of Physics and Astronomy, Johns Hopkins University, Baltimore, MD 21218, USA}

\author[0000-0003-2764-7093]{Masao Sako}
\affiliation{Department of Physics and Astronomy, University of Pennsylvania, Philadelphia, PA 19104, USA}

\author[0000-0002-8538-9195]{Brian Schmidt}
\affiliation{The Research School of Astronomy and Astrophysics, The Australian National University, Canberra, ACT 2611, Australia}

\author[0009-0007-3185-7030]{Xianzhe TZ Tang}
\affiliation{Institute for Astrophysical Research, Boston University, 725 Commonwealth Avenue, Boston, MA 02215, USA}

\author[0000-0002-4283-5159]{Brad~E.~Tucker}
\affiliation{Mt Stromlo Observatory, The Research School of Astronomy and Astrophysics, Australian National University, ACT 2611, Australia}
\affiliation{Australian National Centre for the Public Awareness of Science, Australian National University, ACT 2601, Australia}

\begin{abstract}

Precise measurements of Type Ia supernovae (SNe~Ia) at low redshifts ($z$) serve as one of the most viable keys to unlocking our understanding of cosmic expansion, isotropy, and growth of structure. The Dark Energy Bedrock All-Sky Supernovae (DEBASS) program will deliver the largest uniformly calibrated low-$z$ SN~Ia data set in the southern hemisphere to date. DEBASS utilizes the Dark Energy Camera to image supernovae in conjunction with the Wide-Field Spectrograph (WiFeS) to gather comprehensive host galaxy information. By using the same photometric instrument as both the Dark Energy Survey (DES) and the DECam Local Volume Exploration Survey, DEBASS not only benefits from a robust photometric pipeline and well-calibrated images across the southern sky, but can replace the historic and external low-$z$ samples that were used in the final DES supernova analysis. DEBASS has accumulated more than 400 spectroscopically confirmed SNe~Ia in the redshift range of $0.01<z<0.08$ from 2021 to mid-2025, and, in this paper along with a companion paper \citep{acevedo:inprep-a}, we present an early data release of 77 SNe within the DES footprint to demonstrate the merit and constraining power of the data set. Here, we introduce the DEBASS program, discuss its scientific goals and the advantages it offers for supernova cosmology, and present our initial results. With this early data release, we find a robust median absolute standard deviation of Hubble diagram residuals of $\sim$0.10 mag and an initial measurement of the host-galaxy mass step of $0.06\pm0.04$ mag, both before performing bias corrections. This low scatter shows the promise of a low-$z$ SN~Ia program with a well-calibrated telescope and high signal-to-noise ratio across multiple bands. Data is available on \href{https://github.com/DEBASSCollaboration/DEBASS-DR0.5.git}{GitHub}.

\end{abstract}

\section{Introduction} \label{sec:intro}

Today, cosmology faces a puzzling set of challenges. While the $\Lambda$-Cold Dark Matter ($\Lambda$CDM) model has historically served as the cornerstone of modern cosmology \citep{Betoule2014, Scolnic_2018, Brout2019, Brout2022_a}, explaining the universe’s large-scale structure, cosmic microwave background (CMB), and late-time accelerated expansion, the “Hubble Tension” has presented a stark challenge to the model, persisting for over a decade \citep{Knox2020}. This Hubble Tension problem is illustrated by a significant discrepancy between measurements of the Hubble constant ($H_0$) from late-universe observations like Type Ia supernovae (SNe~Ia; \citealt{Riess2022}) using the distance ladder and $\Lambda$CDM model-based predictions from early-universe data such as the cosmic microwave background (CMB; \citealt{planck2020}). Compounding this issue, recent analyses combining Baryon Acoustic Oscillation data from the Dark Energy Spectroscopic Instrument (DESI; \citealt{Adame_2025}), the Pantheon+ supernova compilation (\citealt{Scolnic2022,Brout2022_a}), Union3 (\citealt{union3}), and the Dark Energy Survey Year 5 supernova analysis (DES-SN5YR; \citealt{Abbott2024}) suggest a preference for a ``thawing" dark energy, deviating from a cosmological constant at $\sim 3 \sigma$ ($w_0, w_a \neq -1, 0$). In both of these cases, the most precise constraints on the late-time universe rely on a shared set of low-redshift (low-$z$) SNe~Ia --- data from older surveys that remain both statistically and systematically limited in many ways. In this paper, we introduce a new low-$z$ survey designed to provide a more robust SN~Ia sample and serve as a foundation for addressing these central tensions in cosmology.

-
The Dark Energy Bedrock All-Sky Supernovae (DEBASS) program is a low-$z$ ($0.001 < z < 0.08$) survey of more than 400 spectroscopically confirmed Type Ia SNe in the southern hemisphere. DEBASS utilizes the same observing instrument (the Dark Energy Camera; \citealt{Flaugher2015}) as the larger and higher redshift ($0.08 < z < 1.2$) DES-SN5YR sample (\citealt{sanchez2024darkenergysurveysupernova}), leveraging DES's characterization of the photometric system and the core of its image processing pipeline. DEBASS will improve upon the samples made by other low-$z$ surveys with exquisite calibration down to 1.8 milli-mag (absolute and all-sky uniformity; \citealt{rykoffdesy6}), well-modeled sample selection, and high signal-to-noise light curves, resulting in improved cosmological parameter constraints. The data described here will ultimately be analyzed in combination with the DES-SN5YR data --- one of the most powerful SN~Ia data sets to date with over 1,500 high redshift SNe~Ia --- resulting in a statistically powerful and systematically robust cosmological analysis with data that is largely independent from Pantheon+. Furthermore, DEBASS will produce pathfinding constraints on the growth rate of structure, $f\sigma_{8}$, in the southern sky and will provide new constraints on our local motion relative to the CMB frame and the impact of local cosmography on measurements of $H_0$ using the southern hemisphere data.

DEBASS coincides with the much-anticipated Zwicky Transient Facility (ZTF) low-$z$ data set \citep{ztf2}. DEBASS relies on ZTF and other all-sky surveys, as discussed later, for discovery and unbiased targeted follow-up with DECam, but distinguishes itself with three primary differences. First, DEBASS targets SNe in the southern hemisphere while ZTF observes in the northern hemisphere. Second, ZTF gathers SN~Ia data in two to three observing bands only (ZTF $g,r,+i$; \citealp{ztfov}), while DEBASS uses all four of DECam's optical bands, $g,~r,~i,~z$. Finally, DEBASS can be combined with the high-redshift DES-SN5YR sample on the same photometric system for systematically-robust cosmological constraints, whereas ZTF will require a cross-calibration effort with external historic high-redshift surveys (similarly to Pantheon+; \citealp{Brout2022}). 

Another more comparable, contemporary low-$z$ survey is the Young Supernova Experiment (YSE; \citealt{Jones2021}). YSE also has a complementary high-redshift survey on the same photometric system (Panoramic Survey Telescope and Rapid Response System (Pan-STARRS) PS1; \citealt{panstarrs}). The PS1 component of YSE is also in the southern hemisphere. For low-$z$ SN cosmology, the all-sky coverage afforded by the combination of DEBASS and YSE will ultimately facilitate incredible statistical constraining power for these volume limited samples.

In this paper, we establish the design and methods of the DEBASS program and present our first results -- referred to as Data Release 0.5 (hereafter DR0.5) -- which demonstrate the merit of the survey to generate state-of-the-art cosmological constraints. Details on survey calibration, simulations and validation, and the first cosmology results are presented in the companion paper \cite{acevedo:inprep-a}. The current paper flows as follows. Section \ref{sec:sdesign} expounds on the motivation for DEBASS, the instruments used to collect the data, the image processing pipeline, the survey strategy used to assemble DEBASS SNe, and the follow-up process. Section \ref{od} covers our observations for all of DEBASS and the processing of the DR0.5 data for the data release given here. Section \ref{res} provides the first results of DEBASS. Finally, in Section \ref{Disc}, we generalize the results of DR0.5 and outline the next steps for DEBASS, including the the near future release of the first 500 SNe which have already been acquired and are currently being processed. DEBASS DR0.5 data has been made available on \href{https://github.com/DEBASSCollaboration/DEBASS-DR0.5.git}{GitHub} and in Table \ref{snedets}.

\section{Survey Design} \label{sec:sdesign}

\subsection{Instruments}
DEBASS utilizes the Dark Energy Camera (DECam) to conduct its photometric follow-up. Originally built for and used by DES to survey a large footprint in the southern sky with both a 5,000 deg$^2$ wide-area survey and 27 deg$^2$ deep drilling field supernova survey,  DECam is mounted on the Blanco 4-meter Telescope at the Cerro Tololo Inter-American Observatory (CTIO). DECam offers a 3 deg$^2$ field of view at a resolution of 0.263 arcseconds per pixel. and is equipped with a set of filters ($grizY$) spanning 4000--10,650 \AA. In addition to DES and DEBASS, DECam has been used extensively for other cosmological surveys (including the DECam Local Volume Exploration Survey, or DELVE; \citealp{DW_2022}), leading to vast and deep southern sky coverage in $griz$ and large overlapping footprints to evaluate photometric calibration uniformity.

A spectroscopic follow-up program runs alongside the DEBASS survey with two key objectives: 1) ensure the spectroscopic classification of every SN in the DEBASS sample as a Type Ia; and 2) measure precise redshifts of each target to produce a Hubble diagram. This spectroscopic follow-up is carried out with the Wide-Field Spectrograph \citep[WiFeS;][]{Dopita2007}, an integral field spectrograph (IFS) mounted on the Australian National University (ANU) 2.3 m Telescope, located at Siding Spring Observatory in Coonabarabran, Australia.
WiFeS is divided into 25 1$\times$38 arcsecond slitlets, which are projected onto 4096$\times$4096 pixel CCDs in each of the red and blue arms of the instrument. By using a 2$\times$ spatial binning, this results in a grid of 1.0$\times$1.0 arcsecond spaxels across the 25$\times$38 arcsecond field of view. Using the B3000 and R3000 gratings along with the RT560 dichroic, we can combine the red and blue arms to produce spectra covering a wavelength range of 3500-9000\AA with a resolution of $R=3000$. Spectroscopic follow-up is done for most of the DEBASS SNe either before or during the early stages of DECam photometry (see Section \ref{bailey}).

\subsection{Target Selection, Exposure Time, and Cadence}\label{targets}
We select DEBASS SNe~Ia from rolling all-sky surveys including the Asteroid Terrestrial-impact Last Alert System (ATLAS; \citealt{Tonry2018}), ZTF, and Pan-STARRS \citep{panstarrs}. These surveys provide a magnitude-limited selection of SNe~Ia (Table \ref{tab:limmag}), ensuring broad coverage of different environments and host galaxy types.
\begin{table}[t!]
\begin{center}
\begin{tabular}{ c c c  }
\hline
Survey & Limiting Mag & Source \\
\hline
 ATLAS & 19.0 ($o$) & \cite{Smith2020} \\
 ZTF & 20.8 ($g$) & \cite{Bellm2018} \\ 
 Pan-STARRS & 23.1 ($i$) & \cite{pslimmag} \\
\hline
\end{tabular}
\caption{$5\sigma$ limiting magnitudes of main DEBASS transient source surveys. \label{tab:limmag}}
\end{center}
\end{table}
Only SNe~Ia that meet specific criteria are selected for follow-up observations: they must be in the Hubble flow ($z > 0.015$, or in a potential Cepheid or Tip of the Red Giant Branch host) but still low redshift ($z < 0.08$); have low Milky Way extinction ($E(B-V)< 0.25$ mag); and be observable by DECam for at least 30 days. This latter criterion ensures that DEBASS is able to capture a complete SN Ia light-curve with an adaptive cadence, as it has a cumulative total of three nights of telescope time each observing semester as a survey within the DECam time-domain collaboration DECam Alliance for Transients (DECAT).

Follow-up for potential DEBASS SNe begins immediately upon detection by one of the aforementioned surveys, whether they are preliminarily classified as SNe~Ia, awaiting SN community classification but with a rising light-curve, or upon DEBASS WiFeS spectroscopic confirmation. In the cases for which final classification suggests a non-Ia or Ia subtype, we abort DECam observations.
We observe each target in all of the $griz$ filters with a default exposure time of 15 seconds per filter. For very bright nearby objects (i.e., $z<0.02$) we adjust the exposure times to avoid saturation. On occasion, DEBASS photometric data is enhanced by observations contributed by other DECam programs like YSE, leading to some SNe being captured in part with 30s or greater exposures. 

With our adaptive cadence, we prioritize more frequent observations around the peak of the target's light-curve when possible (88\% of our DR0.5 data has data before the peak, and 74\% at least 5 days pre-peak),  while post-peak observations are distributed to capture the full decline of the SN's brightness efficiently. This strategy allows for precise measurement of the targets' colors and magnitudes, crucial for accurate distance estimation.

\subsection{Image and light-curve Processing}\label{imgproc}

To process DEBASS images, we employ a data reduction pipeline built on utilities from DES-SN processing \citep{Kessler2015,Morganson2018}. Here we utilize the DES-SN pipeline as adapted for all-sky use by the Dark Energy Survey Gravitational Wave (DES-GW) group \citep{Herner2020a,Herner2020b,thesis} with several modifications to allow for the use of non-uniform and non-aligned public templates. The above references provide specifics on how the pipeline functions; however, we provide here an overview of how we extract photometric data from our raw DECam science images. The images first undergo Single Epoch Processing \citep{Morganson2018}, which performs pixel-level corrections to refine the images and make them science-ready. It then identifies, using Source Extractor \citep{Bertin1996,2011ASPC..442..435B}, any objects within the image and models the point spread function. We crossmatch objects discovered in the image with DELVE in order to calibrate our images and extract zeropoints. DELVE, over the DES footprint, has been shown to agree extremely well with the DES Six-Year Forward Global Calibration
Method catalog \citep{rykoffdesy6}. The companion paper \cite{acevedo:inprep-a} covers our calibration in detail. 

Once the image has completed all the preparatory steps, it undergoes Difference Imaging \citep{Kessler2015}.\footnote{DEBASS Difference Imaging adheres closest to the process outlined in \citep{thesis}, with sum minor adjustments for SN-processing} Difference Imaging takes a given follow-up CCD image and subtracts it from a template image showing the same area of the sky at a time prior to the follow-up image being taken or sufficiently long after the SN's peak to highlight any transients in the area. In DEBASS, we limit our templates to be from at least 30 days prior to the follow-up image or 200+ days after. \begin{figure}[h]
\centering
\includegraphics[width=8cm]{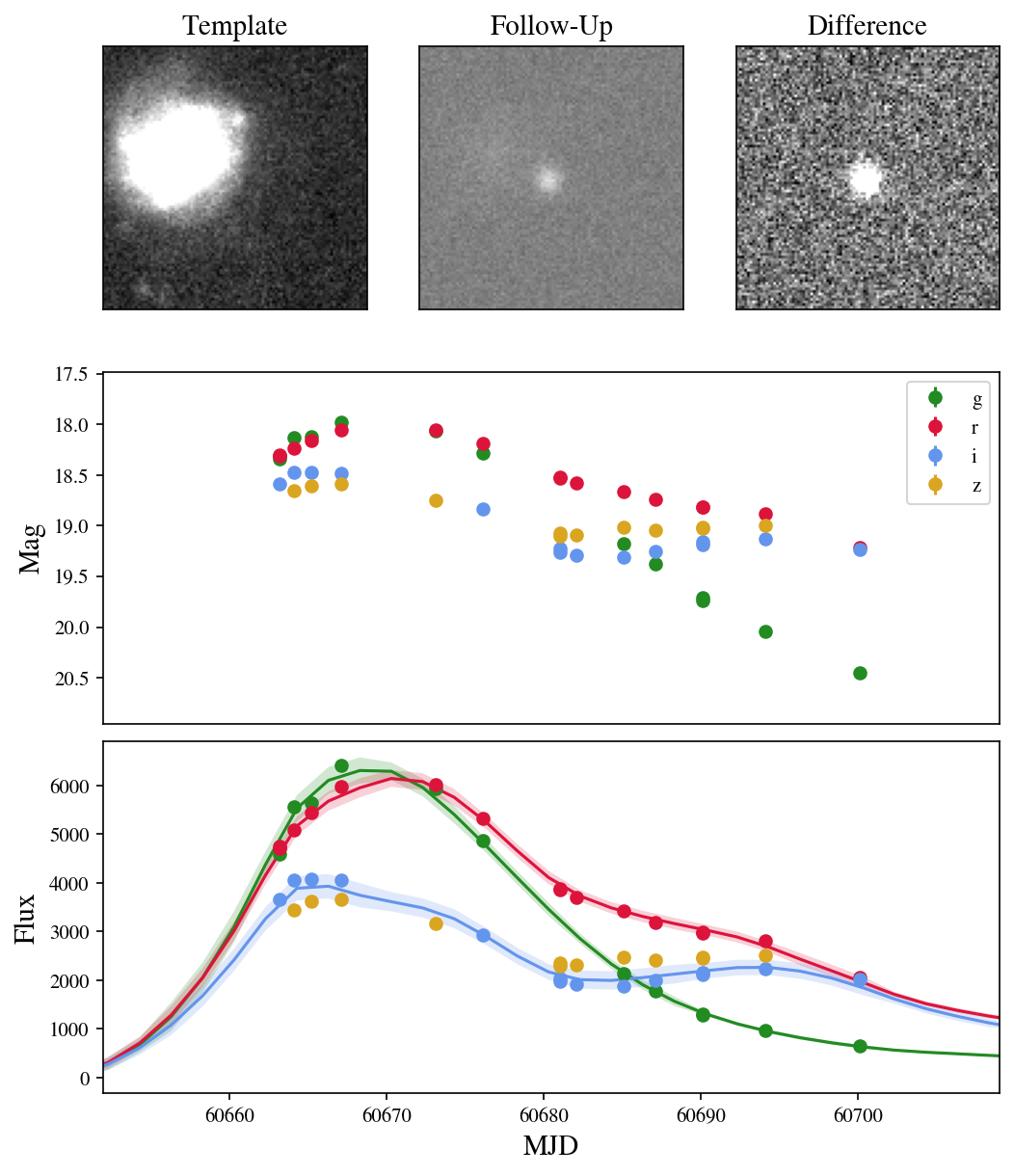}
\caption{Outputs from image processing for DEBASS supernova 2024adzq. \textbf{\textit{Top:}} 51 $\times$ 51 pixel stamps of, from left to right, template, follow-up, and difference images of the supernova. In both the follow-up and the template images, the host galaxy is visible (faint in the shown follow-up image given the short exposure time). \textbf{\textit{Middle:}} Magnitude-space light curve using a 27.5 zeropoint value. \textbf{\textit{Bottom:}} Flux-space light curve with SALT3 light curve fit. 
\label{fig:stamps}}
\end{figure}
\begin{figure*}[ht!]
\centering
\includegraphics{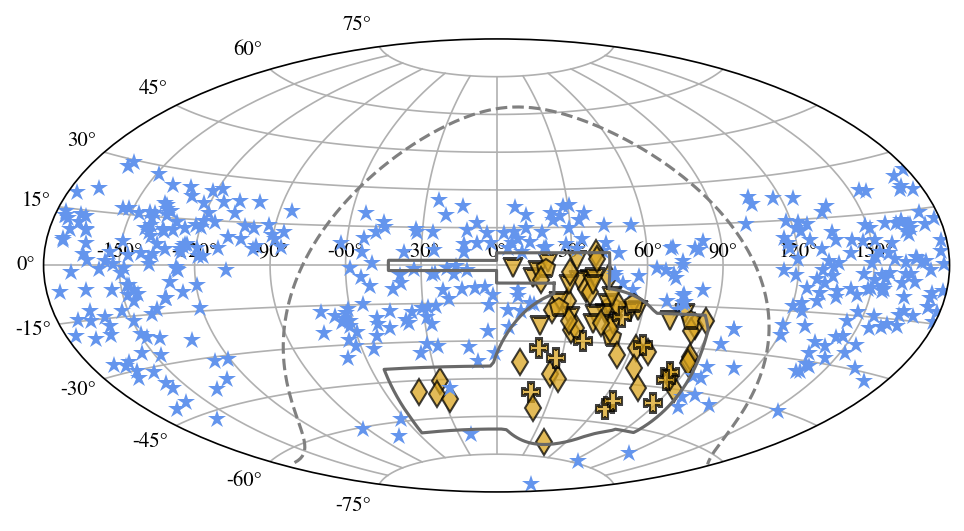}
\caption{All DEBASS SNe at the writing of this paper. Gold makers within the 5,000 deg$^2$ DES footprint (grey outline) designate the 77 SNe that have been processed in the DES wide fields for DR0.5, while the blue stars indicate the remaining DEBASS SNe. The different marker shapes in gold indicate SNe identified by ATLAS (diamond), ZTF (triangle), Pan-STARRS (pentagon), or other (plus) surveys. Table \ref{snesurv} provides an overview of the DEBASS sources. \label{fig:skymap}}
\end{figure*} 

DR0.5 templates have an average exposure time of approximately 100s with an effective time factor of $t_\mathrm{eff}\approx0.66$ across all bands.\footnote{The effective time factor illustrates the quality of the image when considering observing factors including environmental variables. A 100s exposure with a $t_\mathrm{eff}=0.6$ is equivalent to a 60s exposure taken under the same conditions under which the $t_\mathrm{eff}=1.0$ standard was set.} 

Once novel optical transients (real or artifacts of subtraction) are identified in the difference image, again with Source Extractor, we generate 51$\times$51 pixel ``stamps" of the follow-up, template, and difference images centered on the transient object.
Figure \ref{fig:stamps} (Top) shows a real example of these stamps for one of the DEBASS SNe~Ia.

When processing DEBASS optical data, the location of the target SN is already known, avoiding the need for real versus artifact (or non-SN~Ia) classification \citep{Goldstein_2015} in our process.
\begin{center}
\begin{table}[h]
\begin{tabular}{ c c c c }
\hline
ZTF & ATLAS  & Pan-STARRS & Other \\
\hline
 45.4\% & 37.7\%  & 8.45\% & 8.45\% \\ 
\hline
\end{tabular}
\caption{Fractional DEBASS discovery source surveys as reported on TNS. \label{snesurv}}
\end{table}
\end{center}
Given our pipeline's unique ability to detect transients beyond the target candidate --- stemming from its original function for gravitational wave detection --- and the fact that the majority of background real objects in data from this pipeline are likely to be SNe \citep{Morgan2020}, we can study these background objects at a later time. This will allow us to identify supplemental SN~Ia data for a ``photometric" DEBASS sample. With subtraction image stamps at the location of the SNe, we proceed with forced photometry to extract the flux at the DECam-fitted SNe positions, allowing us to construct light curves.

We then perform light-curve fitting using the SuperNova ANAlysis software (SNANA; \citealt{Kessler2009}) jointly with Pippin \citep{Hinton2020} to fit our SN data to the SALT3 SN~Ia model \citep{Kenworthy21} as trained for DES \citep{Taylor_2023}. Section \ref{fit} describes this process in detail. Figure \ref{fig:stamps} shows both an unfitted (in magnitude space, middle graphic) and fitted (in flux space, bottom graphic) light curve for one of our DR0.5 SNe.

\section{Observations and Data}\label{od}

\subsection{Photometric Follow-Up}
As of mid-2025, DEBASS has completed observing for 9 semesters, successfully following up on 558 SNe~Ia. These are depicted in Figure \ref{fig:skymap}. Discovery survey sources for all supernovae (Table \ref{snesurv}) are reported based on the discovery data cataloged in TNS.\footnote{https://www.wis-tns.org} 77 of the DEBASS SNe lie within the DES footprint, providing us with a unique opportunity to reference the DES calibration, simplifying directly comparing and combining with the full DES-SN5YR data set in the future. 

\subsubsection{Observing and Image Processing}\label{obsproc}

On average, our observing cadence yields, for DR0.5, approximately $(12, 13, 12, 9)$ observations in $(g,r,i,z)$ respectively, captures first data $\sim$7 days before peak, and 2 observations per band of pre-peak data. The top right portion of Figure \ref{fig:epochunc} shows the distribution of time between the first observations and peak. While the complete DEBASS data set will contain some SNe at $z>0.08$ due to unknown/imprecise redshifts at the time of DEBASS data collection, we require that all the data from DR0.5 have $z<0.08$. 76\% of our DR0.5 data fall below redshift $z=0.06$, while 56\% fall below $z=0.05$. The mean redshift of our data is 0.044. The table in Appendix A (Table \ref{snedets}) shows the positions, redshifts, and assorted other data of all the DR0.5 SNe.

Image processing results in approximately $(9, 10, 10, 7)$ light-curve data points on average in $(g,r,i,z)$, giving an approximate success rate of 76\%, 78\%, 77\%, and 82\%, per band. Loss of photometric data points can be caused by a variety of reasons, primarily failed image subtraction using HOTPANTS \citep{githubGitHubAcbeckerhotpants}.
\begin{figure*}[ht!]
\centering
\includegraphics[width=16cm]{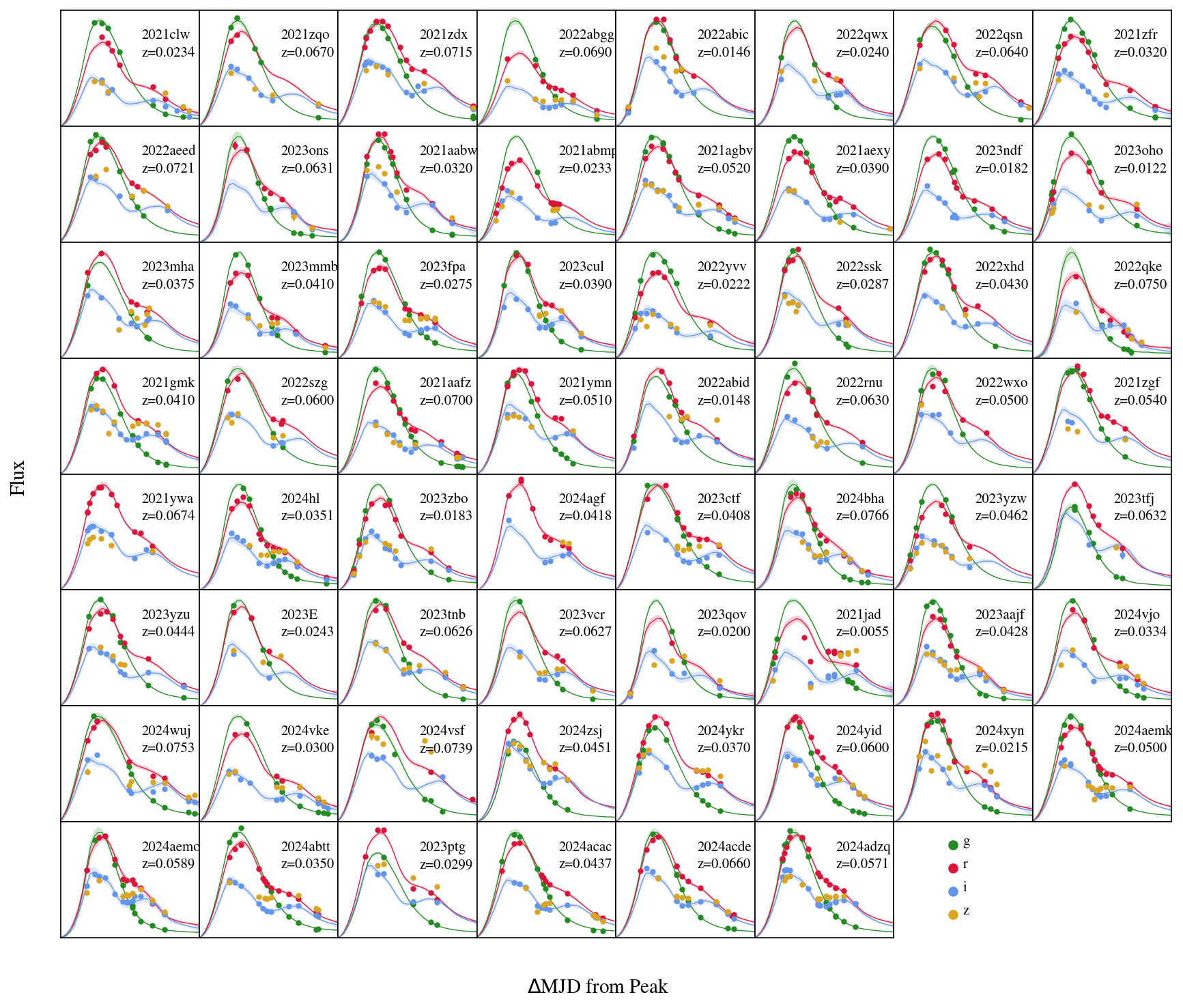}
\caption{All 62 DEBASS DR0.5 supernovae with SNANA fitting and passing cuts in Table \ref{snecuts}. Each subplot shows an individual supernova's light curve across DECam $griz$ bands, shown in forest green, crimson, blue, and gold, respectively, versus $\Delta$MJD from the SN peak observation in days of our fully processed data of the SN. Each subplot is scaled such that x-axis displays $-20 \leq \Delta \mathrm{MJD} \leq 50$ with $\Delta \mathrm{MJD}=0$ being the peak of the given SN and the y-axis ranges from 0 to 1.1 of the peak flux per SN.
\label{fig:snparty}}
\end{figure*}
Finally, we make quality cuts (refer to Table \ref{snecuts} and Section \ref{fit}) to our photometric data including on pixel information and flux and PSF fitting. On average, we retain $(6, 8, 8, 6)$ light-curve data points $(g,r,i,z)$ per SN~Ia, giving an approximate success rate from image processing of 82\%, 85\%, 82\%, and 82\% and from the start of observations of 62\%, 66\%, 64\%, and 67\% per band. Note that these final averages are with respect to the final subset of SNe~Ia subject to cosmology cuts discussed in the next section. Of the 62 SNe surviving cosmology cuts, 59 SNe have observations in 4 optical bands and 3 in only 3. Figure \ref{fig:snparty} shows the fitted light curves of the 62 final SNe using SALT3 DES SN5YR, which does not fit observer frame $z$-band (8250 -- 10150\AA) for the DEBASS sample. Future work is required to extend SALT3 models reliably into the edge of the $z$-band \citep{Pierel2022, Peterson2022,Dai2023}. This does not detract from the analysis power of the DEBASS power compared to other similar surveys.

\subsubsection{Fitted Data}\label{fit}

Image processed data are fitted with SNANA using the same SALT3 model used in DES-SN5YR \citep{mariav2024}. The fitting process yields values for light-curve stretch ($x_1$), color ($c$), peak apparent \textit{B}-band magnitude ($m_B$), light-curve time of peak, and their respective errors.

\begin{figure}[ht!]
\centering
\includegraphics[width=8.5cm]{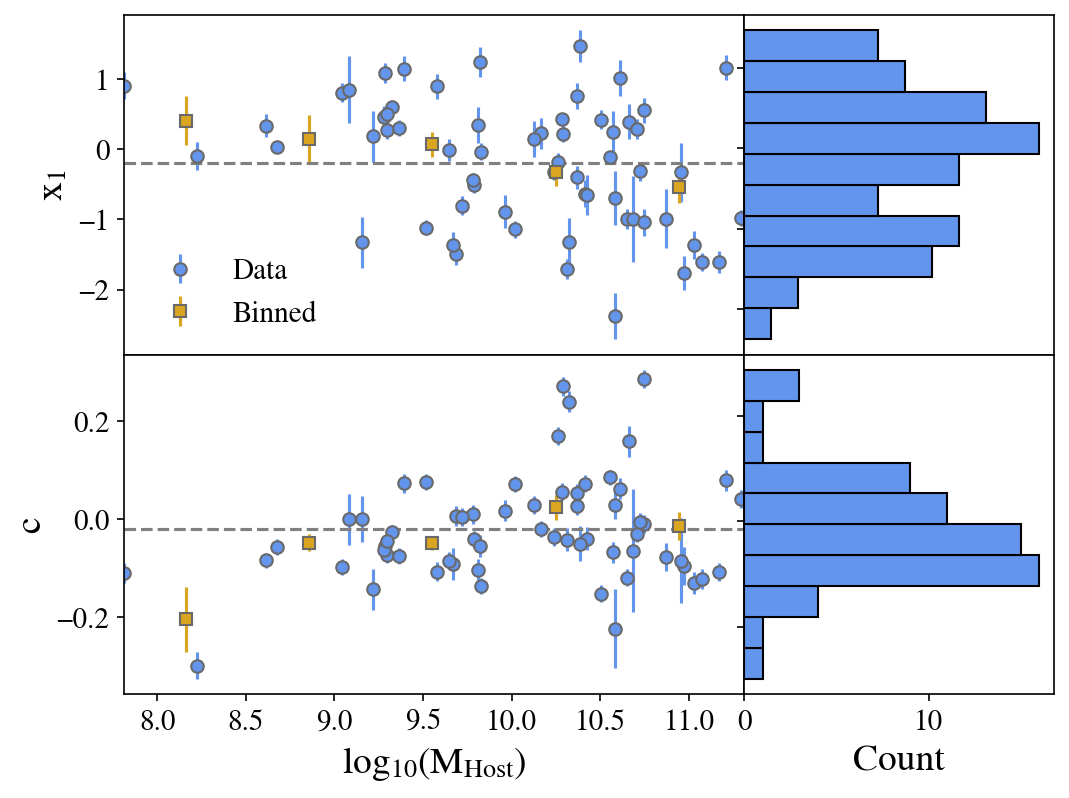}
\caption{The stretch (top) and color (bottom) of the final 62 DR0.5 cosmology SNe~Ia plotted against $\log_{10}{(M_{\mathrm{host}})}$, which we detail in Section \ref{hmass}. The gold points show the data binned by $\log_{10}{(M_{\mathrm{host}})}$. The mean of each of these is shown as a grey dashed line, which is slightly below 0 for both stretch and color. On the right of each graphic, we show the distribution of those data in color and stretch, respectively.
\label{fig:colorstretch}}
\end{figure}

We apply cuts to the final set of SNe~Ia based on the fitting output, adopting the same light-curve quality cuts as DES-SN5YR (\citealp{Brout2019}; see Table \ref{snecuts}) to create our subset of ``cosmology" supernovae. We restrict ourselves to SNe~Ia which have a minimum rest frame observation ($T_\mathrm{{rest_{min}}}$) within 3 days of peak. We filter further such that the $x_1$ uncertainty ($\sigma_{x_{1}}$) must be less than 1.5 and the $c$ uncertainty ($\sigma_{c}$) must be less than 0.15.
We also limit our data such that fitted $\mathrm{MJD}_\mathrm{peak}$ ($\sigma_{\mathrm{MJD}_\mathrm{peak}}$) must be less than 2 days. We place cuts on SN color and stretch, where $c$ must be between $-0.3$ and $0.3$ and $x_1$ must be within $-3$ and $3$ to ensure the fits are within the ranges over which the SALT3 model has been trained. To mitigate Milky Way Extinction, we select only SNe for which $E(B-V)_{MW}<0.2$, although this does not impact our small DR0.5 subset in the DES wide-field. Finally, we apply Chauvenet's Criterion to remove $>3\sigma$ outliers. After these cuts, our data set reduces from 77 to 62 SNe~Ia. Figure \ref{fig:colorstretch} illustrates the $x_1$ and $c$ distributions as completed by SNANA for our final DR0.5 SN~Ia sample. 

\begin{center}
\begin{table}[h]
\begin{tabular}{ c c c }
\hline
Cut & Discarded  & Remaining \\
\hline
 Targeted & --  & 77 \\ 
 SALT3 Fit Converges & 0  & 77 \\ 
$T_\mathrm{{rest_{min}}} \le 3$ & 9 & 68 \\ 
$\sigma_{x_{1}} < 1.5$ & 0 & 68 \\  
$\sigma_{c} < 0.15$ & 0 & 68 \\  
$\sigma_{\mathrm{MJD}_{\mathrm{peak}}} < 2$ & 0 & 68 \\
$-3 < x_{1} <3$ & 1 & 67 \\
$-0.3 < c <0.3$ & 2 & 66 \\
$E(B-V)_{MW}<0.2$ & 0 & 65\\
Chauvenet's Criterion ($3\sigma$) & 3 & 62 \\
 \hline
\end{tabular}
\caption{Cuts applied to fitted data.\label{snecuts}}
\end{table}
\end{center}

\subsection{Spectroscopic Follow-Up}\label{bailey}

In Section \ref{classification}, we outline the methods used to spectroscopically classify DEBASS SNe and describe the process of identifying the host galaxy of each SN in Section \ref{host id}. Our approach to determining the redshifts of each SN in the sample is then described in Section \ref{redshifts}.

\subsubsection{SN Classification}
\label{classification}

As described in Section \ref{targets}, optical follow-up with DECam begins as soon as a potential SN~Ia is identified, often before the transient has a spectroscopic classification. To ensure the DEBASS sample consists entirely of spectroscopically confirmed SNe~Ia and to facilitate future retraining of the SALT3 model, we immediately trigger spectroscopic follow-up with WiFeS on any identified candidates.

\begin{figure*}[ht!]
\centering
\includegraphics[width=18cm]{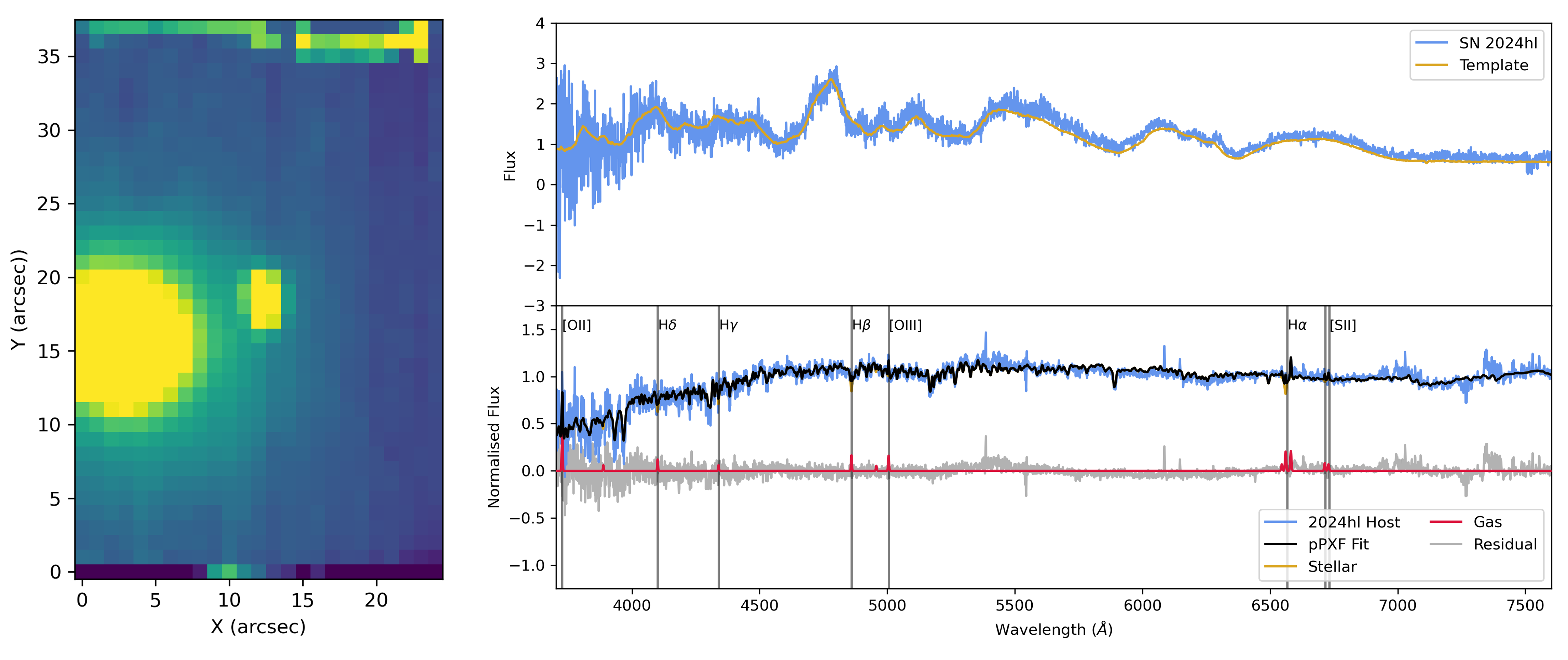}
\caption{\textbf{\textit{Left:}} An example WiFeS datacube from the live follow-up of DEBASS SN 2024hl. Each $1''\times 1''$ spaxel is colored by the mean flux of the corresponding spectrum. The SN can be found at the center of the FoV, to the right of its associated host galaxy. Spaxels at the top and bottom of the reconstructed image are affected by artifacts arising from the ends of the slitlets \textbf{\textit{Right:}} The observed WiFeS spectrum (top) of SN 2024hl is overplotted with the best-fitting template spectrum (gold) from \texttt{SUPERFIT} (SN 1994D, 14 days after light-curve maximum). A \texttt{pPXF} fit to the associated host galaxy is plotted over the observed WiFeS spectrum (bottom). The best-fitting model (black) is broken down into a stellar (gold) and a gaseous (red) component. Key spectral features are highlighted by vertical grey lines.}

\label{fig:merged_spce}
\end{figure*}

We observe using a Nod \& Shuffle approach \citep{Glazebrook_2001} consisting of two 1200s exposures (600s on source, 600s on sky), which enables a high-quality sky subtraction. Data are reduced using the PyWiFeS pipeline \citep{Childress2013}, producing integral-field datacubes (see Figure \ref{fig:merged_spce}, left). From these datacubes, we extract two spectra: one from the transient itself for the purpose of classification and model training; and a second spectrum from a potential host galaxy in the field of view, in order to obtain a host redshift (see Section \ref{redshifts}). 

We classify the observed targets using the \texttt{superfit} program \citep{Howell2005}, which utilizes template spectra of transients at a variety of light-curve phases. \texttt{superfit} also contains a set of template galaxy spectra in order to perform a host galaxy subtraction and reduce any effects of host contamination in our observed data. \texttt{superfit} sequentially compares the observed spectrum to templates in the rest-frame, returning a list of potential matches sorted by their reduced $\chi^2$ value. In most cases, a redshift is pre-determined from our extracted host spectrum, but in the case when this is not possible, \texttt{superfit} can sample a defined redshift range to give an estimate of the best-fitting redshift. These redshifts are then determined more precisely from an additional dedicated host galaxy follow-up program (see Section \ref{hosts}). An example WiFeS spectrum of DEBASS SN 2024hl is shown in Figure \ref{fig:merged_spce}, right, along with the best-fitting template from \texttt{superfit}.

\subsubsection{Host Identification}\label{host id}
Host galaxies are assigned using the directional light radius method (DLR; \citealt{Sullivan_2006, Gupta_2016}). The matching minimizes the dimensionless variable, $d_{\mathrm{DLR}}$, the ratio between the angular separation between the SN and the center of a galaxy, and the elliptical radius of the galaxy in the direction of the SN. This method is highly sensitive to parameter choices, particularly the detection threshold, as only pixels with values exceeding this threshold are considered. Following \cite{Wiseman2020}, we adopt a low detection threshold of 1.5, allowing small and faint objects to be detected. As with \cite{Sako2018} and \cite{ Qu2024}, we use a limit of $d_{\mathrm{DLR}} \leq 4$, such that any SN with no galaxy within this threshold is considered to be ``hostless". We note that there are no hostless SNe~Ia in DEBASS DR0.5.

\subsubsection{Redshifts}\label{redshifts}
We derive spectroscopic redshifts from the host galaxies of each SN, rather than from the SNe themselves, as SN-derived redshifts are less precise, non-repeatable measurements that have been shown to induce biases in the determination of cosmological parameters \citep{Steinhardt2020}. Preliminary redshifts are taken from the WiFeS follow-up of the live SN candidates, by extracting a region of the host galaxy that is displaced from the SN position. This ensures that we minimize the impact of any SN contamination. All DEBASS host galaxies are followed up with WiFeS after their respective SNe have faded (see Section \ref{host follow-up}), and from these higher quality host spectra we obtain a more precise redshift that replaces the preliminary SN-contaminated measurement. 

When measuring redshifts from our observed spectra, we utilize the \texttt{MARZ} tool \citep{Hinton2016}, which is the same software used by DES-SN for this purpose \citep{Lidman2020}. From these observed redshifts, we apply corrections to account for different systematic impacts following the approach of \cite{Carr_2022}.

We first account for the Earth's rotation and orbit around the Sun by converting our redshifts to the heliocentric frame. We do this using the \texttt{astropy} module \citep{2013A&A...558A..33A} to determine a velocity correction factor $v_\text{hel}$ for a given time of observation and sky position. This heliocentric redshift $z_\text{hel}$ is given by
\begin{equation}\label{z_hel}
    1 + z_\text{hel} = (1 + z_\text{obs})(1 + v_\text{hel}/c),
\end{equation}
where $z_\text{obs}$ is the observed redshift from \texttt{MARZ}. Since $v_\text{hel}\ll c$ (of order 10 km/s), we apply a non-relativistic correction factor in the second term. 

We then correct for the peculiar motion of the Sun relative to the CMB, which is projected onto an observed target by 
\begin{equation}\label{v_sun}
    v_\text{Sun} = v_\text{Sun}^\text{max}\text{cos}(\alpha),
\end{equation} where $v_\text{Sun}^\text{max}$ is the velocity of the Sun in the direction of the CMB dipole, and $\alpha$ is the angle between the positions of the dipole and the observed object. We adopt the CMB dipole measurements from the Planck Collaboration \citep{planck2020}, giving a velocity $v_\text{Sun}^\text{max}=369.82\pm0.11\ \text{km}\ \text{s}^{-1}$ in the direction $(l,\ b) = (264.021^\circ \pm 0.011^\circ,\  48.253^\circ \pm 0.005^\circ)$. We use the relativistic approach to determining the redshift contribution $z_\text{Sun}$, given by
\begin{equation}\label{z_sun}
    1 + z_\text{Sun} = \sqrt{\frac{1-v_\text{Sun}/c}{1+v_\text{Sun}/c}},
\end{equation}
allowing us to determine the dipole-corrected redshift $z_\text{CMB}$ of each SN in the sample:
\begin{equation}\label{z_cmb}
    1 + z_\text{CMB} = \frac{1 + z_\text{hel}}{1 + z_\text{Sun}}.
\end{equation}

Finally, we apply a correction for the peculiar velocities (PVs) of the SN host galaxies themselves. In order to perform precise cosmology using low-redshift SNe~Ia, we must distinguish the components of redshift from peculiar motion and universe expansion. We obtain our PVs using the methods documented in \cite{Peterson2022}, reconstructing a velocity field from a density field using primarily 2M++ \citep{Lavaux2011, Carrick2015} and SDSS \citep{Albareti2017}. 
With these PVs, we can obtain a correction factor $z_\text{PV}$ which is applied in the same manner as above \cite{Carr_2022}:
\begin{equation}\label{z_HD}
    1 + z_\text{HD} = \frac{1 + z_\text{CMB}}{1 + z_\mathrm{PV}},
\end{equation}
where $z_\text{HD}$ is the Hubble diagram redshift that is useful for cosmological analysis with this data set. In the data release, we also provide the uncorrected $z_\text{CMB}$ for use in PV analyses.

\subsection{Host Properties}\label{hosts}

The low-redshift nature of DEBASS provides a valuable opportunity to investigate the impact of host galaxies on SN~Ia light curves and their resulting distance measurements. This is particularly motivated by the existence of a ``luminosity step" seen when comparing SN Hubble residuals with different host galaxy properties. The correlation between Hubble residuals and host galaxy stellar mass (see Section \ref{massstep}), whereby more massive galaxies host systematically brighter SNe~Ia after standard light-curve corrections, has been known for over a decade \citep{Sullivan2010, Childress2013, Smith2020, Kelsey2020}; however, significant correlations with other host galaxy properties have been demonstrated. These include the gas-phase metallicity \citep{Pan2014, Campbell2016, Galbany2022}, specific star formation rate \citep[sSFR;][]{Childress2013, Rigault2020, Martin2024, Dixon2025}, stellar age \citep{Rose2021, Wiseman2023}, and rest-frame color \citep{Jones2018, Roman2018, Kelsey2021}. We determine photometric and spectroscopic properties of every host galaxy in the DEBASS sample in order to complete a rigorous analysis of host systematics for cosmology with DEBASS.

\subsubsection{Host Photometry and Global Properties}\label{hmass}

Here we briefly describe the methods used to obtain global photometry of each SN~Ia host galaxy. We use images from public surveys, including Pan-STARRS Data Release 1 \citep{panstarrs}, DES \citep{DES}, and the Sloan Digital Sky Survey (SDSS; \citealp{SDSS}), which are made available in \texttt{Hostphot} \citep{Hostphot} with broadband coverage in $grizy$, $grizY$, and $ugriz$, respectively. These bands are well-suited for estimating global properties.

We use \texttt{Hostphot}\footnote{We used \texttt{v2.8.0}.} to estimate broad-band magnitudes. We first define the aperture to be used for all filters using a co-added image created from the $riz$ filters. Other sources in the co-added image (e.g.,~bright foreground stars near the host galaxy) are masked out. Masks are extracted and subsequently applied to other single filter images to ensure consistent photometry.

We estimate magnitudes using \texttt{Hostphot}'s global photometric method, ensuring the same elliptical aperture is applied across all filters. This aperture is determined from the co-added image using the optimized Kron-radius scale \citep{Kron1980}. We choose Kron fluxes over isophotal fluxes, as the latter tend to exclude low-surface-brightness flux in galaxy outskirts, biasing host galaxy mass estimates downward. The resulting fluxes in each filter form a coarse spectral energy distribution (SED) for each galaxy.

The fluxes are fitted with synthetic stellar population models. The method for host mass estimation is identical to the method used for the DES-SN5YR analysis \citep{Vincenzi2024} and is based on the work of \cite{Sullivan2010}. Prior to fitting, all photometric measurements are corrected for Milky Way extinction using the re-calibrated dust maps of \cite{Schlafly_Finkbeiner} and the extinction law from \cite{Fitzpatrick_1999}. We perform SED fitting by comparing the multi-band photometric observations of each host galaxy to a grid of synthetic SED models generated using the \texttt{PEGASE2} \citep{PEGASE}, which provides model spectra based on varying assumptions about star formation histories, metallicity, and stellar evolution. We adopt a \cite{Kroupa_2001} initial mass function and assume a Flat $\Lambda \mathrm{CDM}$ cosmology with $H_0 = 70$, $\Omega_{M}=0.3$, and $\Omega_\Lambda=0.7$. From this SED fitting we obtain host-galaxy mass and rest-frame colors. Final stellar masses are expressed in the form of $\log_{10}(M_{Host}/M_{\odot})$. 

We focus on consistency in mass determination across redshift, which has been emphasized in \cite{vincenzi2025comparingdessn5yrpantheonsn}. We re-compute host galaxy masses for the Foundation sample and compare to common low-$z$ hosts in DES-SN5YR using the same pipeline applied to DEBASS sample. The resulting Pearson correlation coefficient of 0.926 and a median offset of $-0.029$ dex indicates strong agreement across samples, thereby validating the reliability of our host mass estimates for DEBASS sample. 

As in \cite{Kelsey2023}, we also compute rest frame $(U-R)$ color\footnote{By adopting the Bessell transmission functions for both bands, we quote magnitudes in Vega system.}. We choose this color because it covers a broad spectral range and is the most widely used color in literature. As of mid-2025, stellar mass, rest-frame $(U - R)$ color, and directional light radius have been measured for approximately 400 DEBASS host galaxies, including all those within the DES footprint presented here in DR0.5.

\subsubsection{Host Spectroscopy}\label{host follow-up}

Many previous studies that have investigated spectroscopic properties of SN~Ia host galaxies have focused on higher redshifts probed by large SN~Ia surveys such as DES \citep[e.g.,][]{Galbany2022, Dixon2025}. While an increased sample size is provided by these data sets, the signal-to-noise ratio (S/N) of the individual host spectra are often too poor, meaning that these spectra must be stacked or their properties binned in order to reduce the uncertainties in the measurements of each property, resulting in sets of $\sim$10 datapoints to use for a correlation analysis. The low-redshift nature of DEBASS means that we can obtain host spectra with a high enough S/N to measure properties of individual host galaxies, increasing the sample size used for correlation analysis by at least an order of magnitude. 

We are in the process of completing a dedicated spectroscopic follow-up of every DEBASS host galaxy with WiFeS on the ANU 2.3m telescope. These galaxies are observed at least 90 days after the light-curve peak of their respective SNe to minimize contamination from the SN. We use a similar observational setup to the SN classification program, utilizing the B3000 and R3000 filters to give spectra covering 3500-9000\AA with a resolution of $R = 3000$. We use the same Nod \& Shuffle approach, this time completing three 1200s exposures providing a total of 1800s on source and 1800s on sky. This gives us an average S/N$\sim$10 across our sample. Since WiFeS is an IFU spectrograph, if the position of the SN is also within the WiFeS field of view for a given host galaxy observation, then we have the opportunity to investigate both the global properties of the host galaxies, as well as the environments local to the detected SNe. This local analysis is motivated by recent studies that find strengthened Hubble residual correlations when looking at properties of the local environment as opposed to global host galaxy properties \citep{Roman2018,Rigault2020,Rose2021}. To date, we have observed over 200 of the DEBASS host galaxies, including the hosts of all SNe overlapping the DES footprint that are used in this paper.

Once we have extracted our galaxy spectra, we perform spectral fitting to derive host properties. This is completed with the Penalized PiXel-Fitting module \citep[pPXF;][]{Cappellari2004}, which combines single stellar populations (SSPs) from the Extended-MILES spectral library \citep{Vazdekis2016} in order to model an observed galaxy spectrum. From these \texttt{pPXF} fits, we can extract the stellar and gaseous components of the model individually, and measure key host properties such as the mean stellar age and metallicity, the mass to light ratio, and the equivalent widths of numerous emission features. 
An example fit to one of our observed host galaxies are shown in Figure \ref{fig:merged_spce}, right.

\section{Results}\label{res}
\subsection{Initial Hubble Diagram}
\begin{figure*}[ht!]
\centering
\includegraphics[width=16cm]{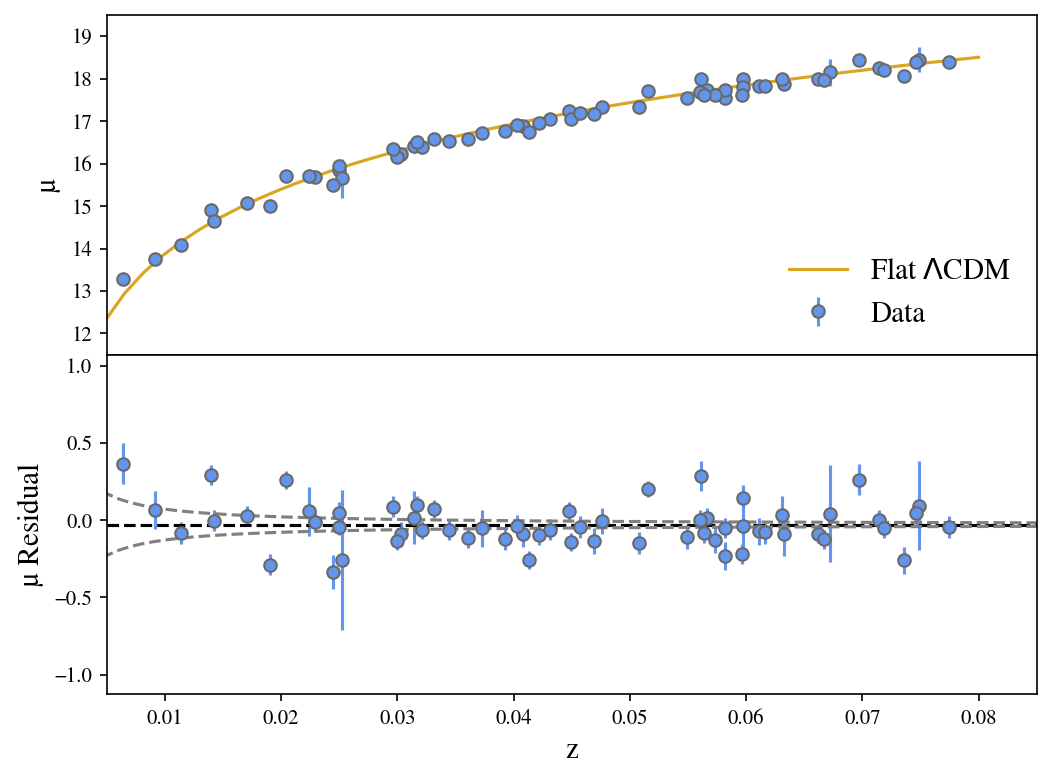}
\caption{\textbf{\textit{Top:}} Hubble diagram using the 62 fitted DEBASS DR0.5 SNe plotted against the ``best fit" cosmology (gold). \textbf{\textit{Bottom:}} Residuals between the computed distance moduli and model. The dashed black line indicates the mean of the residuals while the dashed grey lines account for errors from peculiar velocities of 300 km/s across the data redshift range. Model fitting contributes predominately to our uncertainties.}
\label{fig:hubble}
\end{figure*}
We derive a distance modulus $\mu$ using a modified version of the Tripp Equation \citep{1998A&A...331..815T}
\begin{equation}
    \mu = m_B + \alpha x_1 + \beta c - \mathcal{M},
    \label{tripp}
\end{equation}
populating $m_B$, $x_1$, and $c$ from our SALT3 fitting results for each SN~Ia. We use SNANA's \texttt{SALT2mu} functionality to derive $\alpha$ and $\beta$ and compute the mass step (see Section \ref{massstep}). We find $\alpha=0.17\pm0.02$, $\beta = 2.78\pm0.17$ and note that these values are computed without bias corrections. 
Using \texttt{SALT2mu}, we find a residual scatter ($\sigma_{\mathrm{int}}$) in the Hubble Diagram of $\sigma_{\mathrm{int}} = 0.12$ mag. 

Using a Flat $\Lambda$CDM cosmology with blinded $\Omega_{m}$ found by minimizing the $\chi^2$ of the cosmology fit, we can derive a $\mu_{\mathrm{model}}$ in order to compute Hubble residuals ($\mu - \mu_{\mathrm{model}}$) and construct a Hubble diagram (Figure \ref{fig:hubble}).
We compute Hubble residual scatter using the robust median absolute standard deviation (RSD; \citealt{Hoaglin00})

\begin{multline}
    \mathrm{RSD} = 
     1.48 \times \mathrm{median}(|(\mu - \mu_{\Lambda \mathrm{CDM}}) \\ - \mathrm{median}(\mu - \mu_{\Lambda \mathrm{CDM}})|)\,.
\end{multline}
Using only DR0.5's 62 cosmology SNe that have $z>0.02$, we compute an RSD of $0.10$ mag. Note that we report this scatter without bias correction. See \cite{acevedo:inprep-a} for our bias-corrected and simulation analyses. The RSD Hubble scatter is smaller than $\sigma_{\mathrm{int}}$ as it is less sensitive to outliers in the data. The impact of residuals can also be seen when comparing the standard deviation of the Hubble residuals, $\sigma_{\mathrm{std}} = 0.12$ mag, which matches the $\sigma_{\mathrm{int}}$.

\begin{figure}[h!]
\centering
\includegraphics[width=8.5cm]{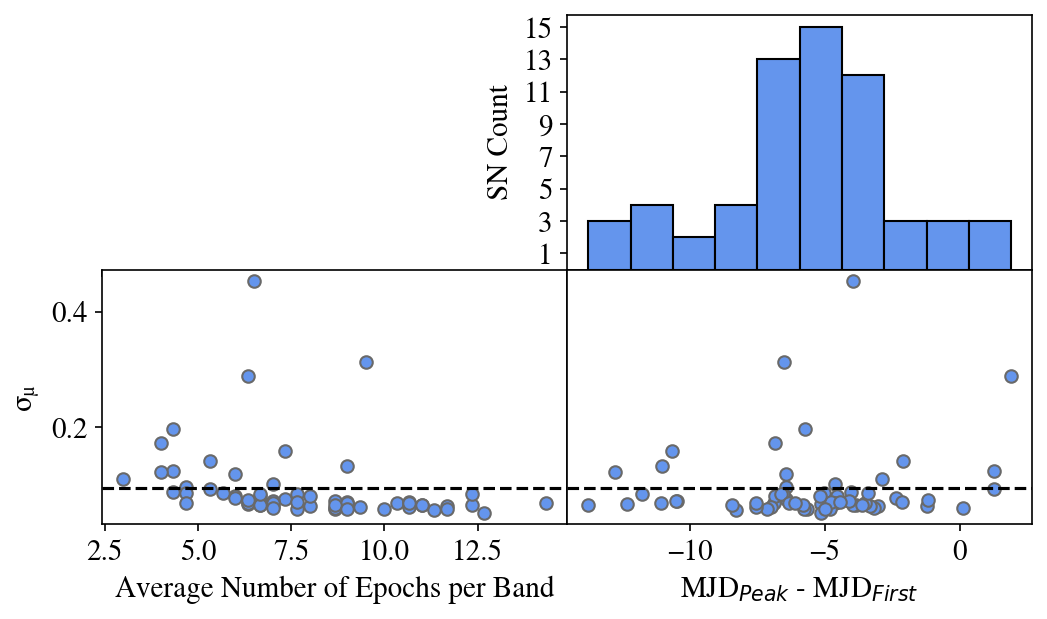}
\caption{Uncertainty on the computed distance moduli for the 62 fitted SN~Ia versus the average number of observations (epochs) imaged per band for the SN (left) and the $\Delta$MJD between the fitted peak MJD and the first observation (right). The dashed black line shows the mean distance modulus uncertainty. Above the graphic on the right, we also show the distribution of $\Delta$MJD between the fitted peak MJD and the first observation, illustrating the success of the DEBASS observing strategy to capture SNe Ia as they rise in brightness.}
\label{fig:epochunc}
\end{figure}

The bottom two panels of Figure \ref{fig:epochunc} show a budget of the uncertainty on our distance moduli by (left) the number of observations per band of each SN and (right) how early we began taking data. We do not include $\sigma_{\mathrm{int}}$ in our calculations of uncertainty on the distance moduli, which is reported only as error from the Tripp Equation (Equation \ref{tripp}).

\subsection{Mass Step}\label{massstep}
\begin{figure}[h!]
\centering
\includegraphics[width=8.5cm]{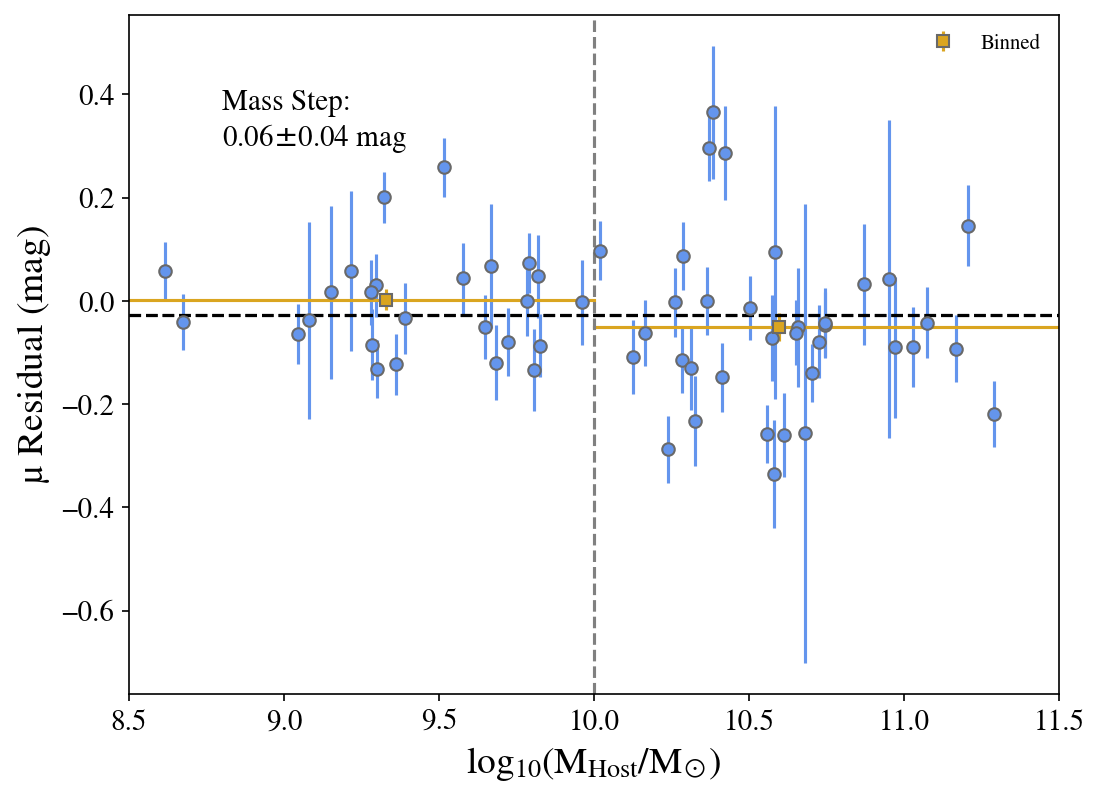}
\caption{DEBASS DR0.5 Hubble residuals (without a mass step correction applied) plotted against the $\log_{10}(M_\mathrm{Host}/M_{\odot} )$ of their host galaxies. There is weak evidence for a step at $\log_{10}(M_\mathrm{Host}/M_{\odot} ) = 10$ which could indicate a correlation between the residuals  and the host mass.
\label{fig:hubble_mass}}
\end{figure}

By comparing the distance modulus residuals to their respective host-galaxy masses, we expect to see evidence of a mass step at $\log_{10}(M_\mathrm{Host}/M_{\odot}) = 10$, a correction factor historically found for luminosity standardization \citep{Betoule2014,Smith2020,Johansson2021,Chung2023} and potentially the result of host dust properties \citep{Brout2021,popovic2022pantheonanalysisforwardmodelingdust, Wiseman2023,popovic2024modellingimpacthostgalaxy}.  Using SALT2mu, we find for the DR0.5 sample a mass step value of $\gamma =0.063\pm0.038$ mag. Figure \ref{fig:hubble_mass} shows the mass step using our manually computed Hubble residuals. Table \ref{values} shows in one location the main numeric results of DR0.5 in this paper.

\begin{center}
\begin{table}[h]
\begin{tabular}{ c c c c c c}
\hline
RSD & $\sigma_{\mathrm{std}}$  & $\sigma_{int}$ & $\alpha$ & $\beta$ & $\gamma$\\
\hline
 0.10 & 0.12  & 0.12 & 0.17 & 2.78 & 0.063
\\ 
\hline
\end{tabular}
\caption{The RSD, $\sigma_\mathrm{std}$ and $\sigma_{int}$ residuals and $\alpha$, $\beta$, and $\gamma$ values from the Tripp Equation (here reported without error) for the DR0.5 data set as found without bias correction. $\sigma_{int}, \,\alpha$, and$\beta$ are all found using SALT2mu with the mass step correction given by $\gamma$. The RSD and $\sigma_\mathrm{std}$ are found using Hubble residuals manually computed from Equation \ref{tripp} using the $\alpha$, and$\beta$ above. \label{values}}
\end{table}
\end{center}

\subsection{Comparisons to Other SN~Ia Surveys}
In order to assess the performance of the DEBASS program and to emphasize both its similarities and differences to other similar low-$z$ surveys, we compare our DR0.5 results to those of Foundation \citep{Foley2017} and ZTF \citep{Dhawan_2021,Ginolin_2025}.

\begin{figure}[h!]
\centering
\includegraphics[width=8.5cm]{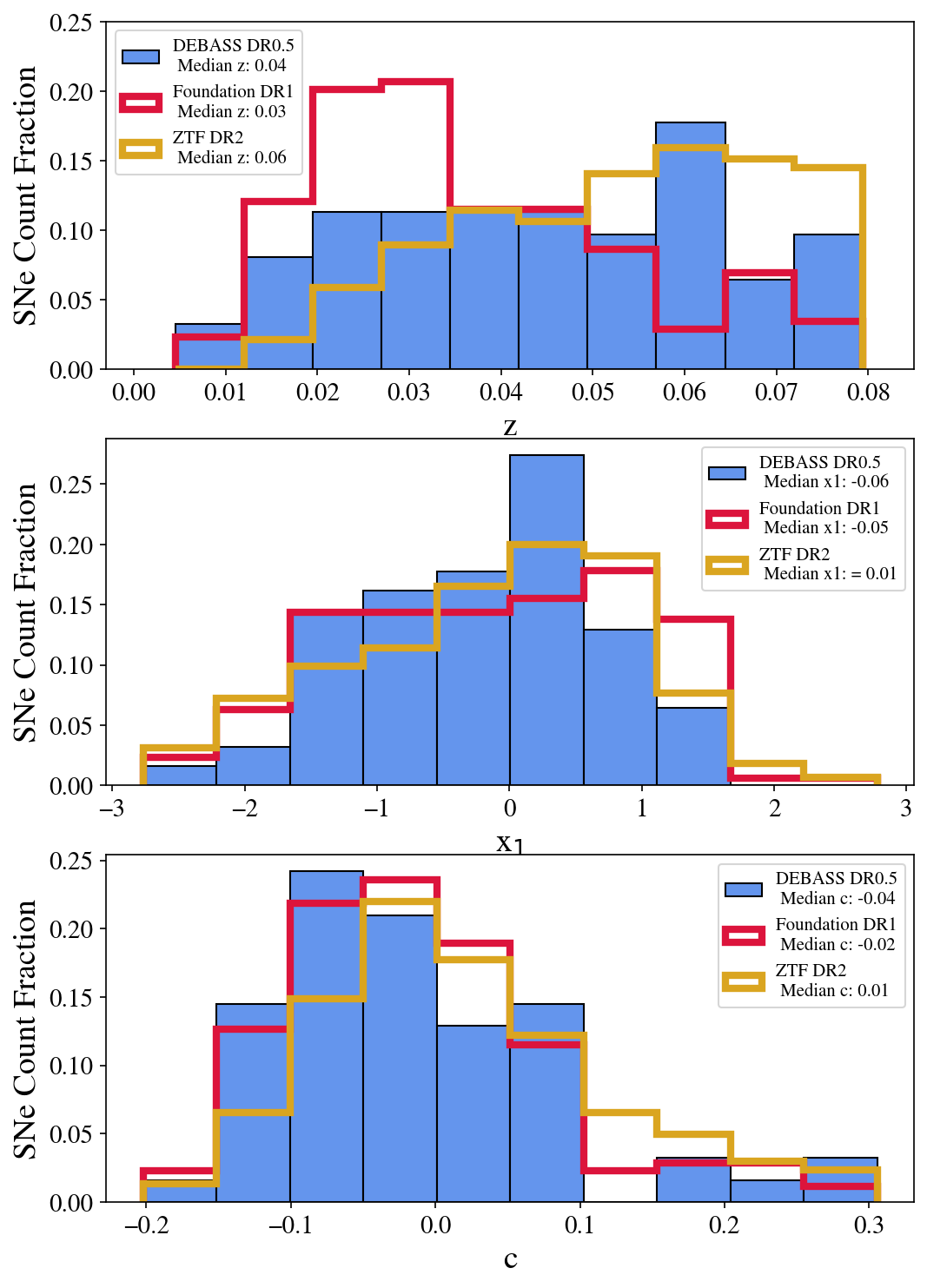}
\caption{Distributions of DEBASS DR0.5 (blue) redshift, stretch, and color values (top, middle, and bottom panels, respectively) compared to low-$z$ surveys Foundation (crimson) and ZTF (gold). 
\label{fig:sncomps}}
\end{figure}

Figure \ref{fig:sncomps} depicts a comparison of the redshift, stretch, and color distributions of these three surveys. The median redshift of the DEBASS DR0.5 falls slightly greater than that of Foundation (0.038) at 0.047. Interestingly there is a deficit of large $x_1$ in the DR0.5 sample compared to Foundation and ZTF; this is discussed in Section \ref{Disc}. The DEBASS DR0.5 color distribution is slightly bluer than that of Foundation and ZTF.

\begin{figure}[h!]
\centering
\includegraphics[width=8.5cm]{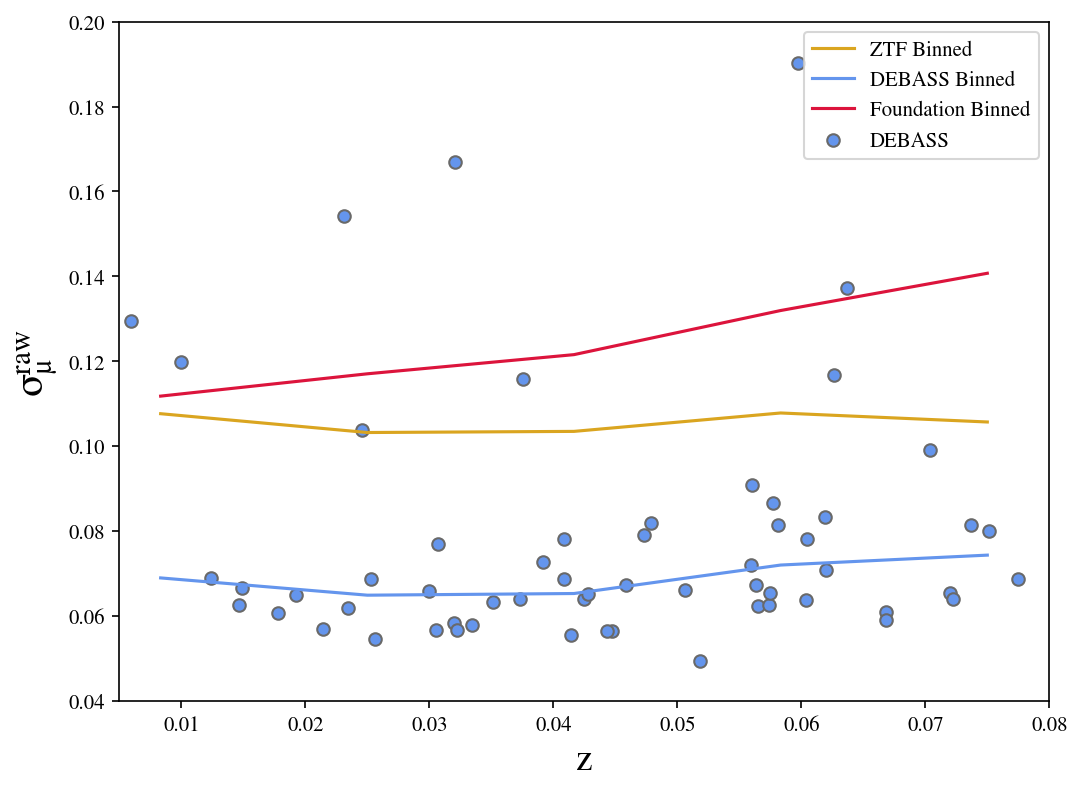}
\caption{Uncertainty on the distance moduli versus redshift for
DEBASS (blue), Foundation (crimson), and ZTF (gold). We compute the raw distance modulous errors (no intrisic scatter floor) for each survey using the Tripp equation and the $\alpha$ and $\beta$ values taken from \cite{mariav2024} for consistency of comparison, although using the values presented in this paper provides no appreciable difference to content of the figure. The solid lines show the binned medians of each survey's uncertainties, while the points show the computed values for individual DEBASS SNe.
\label{fig:fig16}}
\end{figure}

We can also compare the $\sigma_{\mathrm{int}}$ and Hubble residual scatters for each of these programs, noting again, that distances reported here lack bias correction that will result in more accurate cosmology. Foundation reports an intrinsic scatter of $\sigma_\mathrm{int} = 0.111$ mag and weighted root mean square (RMS) of $0.136$ mag \citep{Foley2017} and is consistent with our result. ZTF reports a scatter (using only SNe~Ia with redshifts not derived from SN-features) and peculiar velocity contributions of $0.145$ mag and an overall scatter, computed via standard deviation, of $\sigma = 0.187$  mag \citep{ztfother}, larger than both DR0.5 and Foundation.
Figure \ref{fig:fig16} shows the computed uncertainties of each survey's distance moduli as calculated directly from the error propagation of the Tripp Equation (Equation \ref{tripp}), without the inclusion of an intrinsic scatter floor (denoted $\sigma_{\mu}^\mathrm{raw}$).  To facilitate this comparison of the three surveys, we apply the cosmology cuts listed in Table \ref{snecuts} to both ZTF (without the $\sigma \mathrm{MJD}_{\mathrm{peak}}$ cut as peak data was not provided for ZTF) and Foundation. 

Lastly, we can compare the reported global mass steps found in each SN sample. For DEBASS DR0.5 we find $0.063\pm0.038$ mag. ZTF finds $0.145\pm0.021$ mag for the same analysis method (global step, joint fit of nuisance parameters, single $\alpha$).

\section{Discussion}\label{Disc}
In order to further exemplify the merit of the DEBASS data set, we can contextualize our DR0.5 cosmology data set with comparable low-$z$ surveys, including Foundation and ZTF. Figure \ref{fig:sncomps} does this for redshift, color, and stretch, where data is available in the external surveys. 
DEBASS DR0.5 SNe Ia are, on average, bluer and faster than SNe Ia from both Foundation and ZTF, which might imply that our sample is less dusty than both. This could account for the lower value of total RSD scatter compared to what is seen in previous surveys. Alternately, our tendency towards blue color could be attributed to selection bias in our transient source surveys.

Bluer and lower $x_1$/higher mass distributions have been found in \cite{Brout2022} to give rise to lower scatter in the Hubble diagram. They determined a theoretical minimum Hubble Residual scatter of 0.08 mag, which our scatter approaches. With the full DEBASS sample, our proximity to this minimum value may lessen; however, the current reported value demonstrates the full capabilities of our survey.

Photometric processing and analysis is ongoing for the currently available DEBASS SNe, and will be supplemented with additional targets over time. DEBASS has finished observing 542 SNe~Ia and is currently actively observing an additional 16 (total of 558). These supernovae inhabit $0.001<z<0.14$. In a restricted redshift range of $0.01<z<0.08$, we immediately reduce our full sample to 458 SNe~Ia. We have at least one WiFeS spectrum of the live SN for 209 SNe, a WiFeS host spectrum for 323 of the SNe, and computed masses for 249 SNe (and counting). 
\begin{figure}[ht!]
\centering
\includegraphics[width=8.5cm]{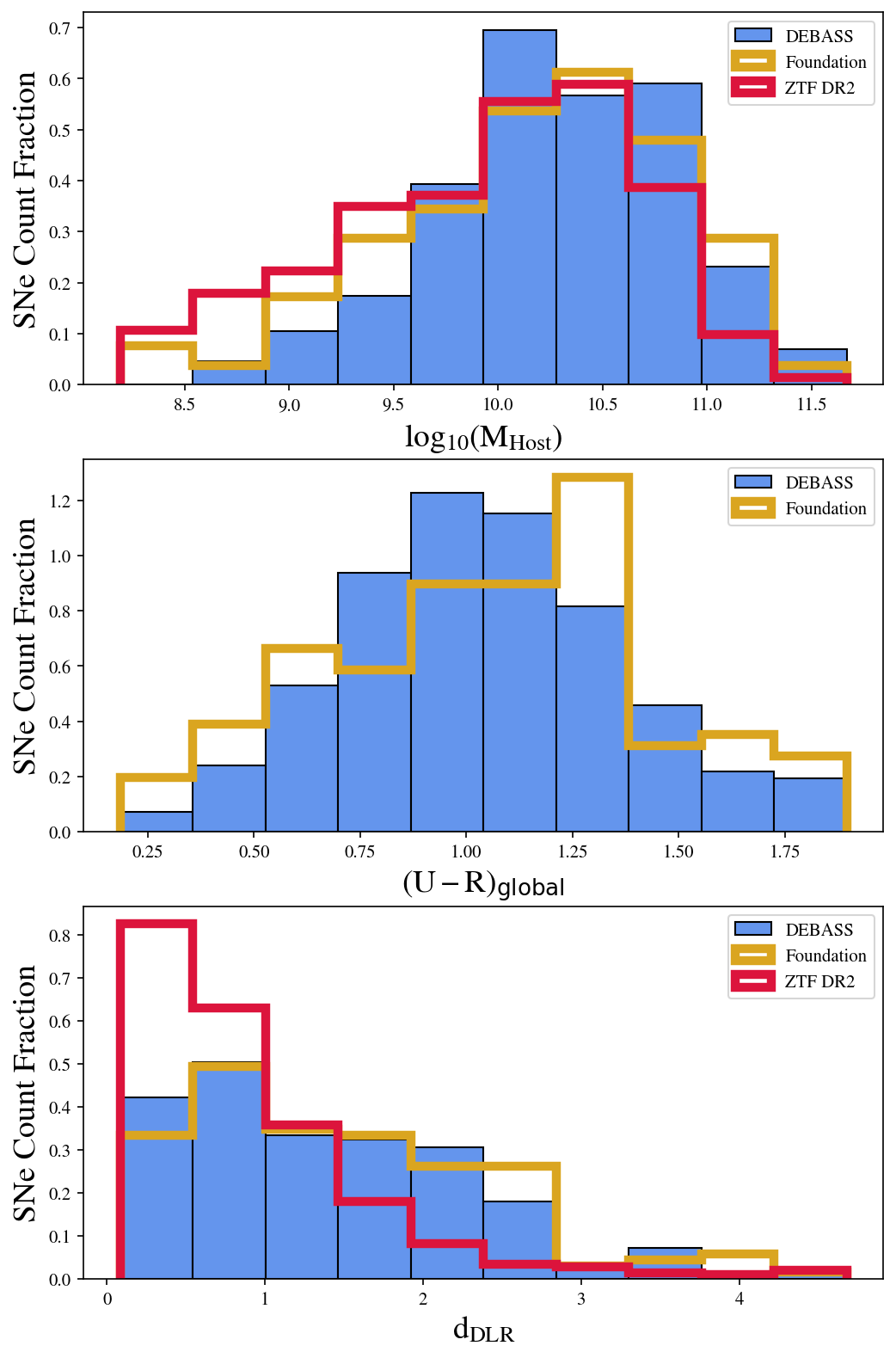}
\caption{All available DEBASS host data compared to other leading low-$z$ data sets where the relative separation of host from SN is $d_{\mathrm{DLR}}<5$. \textit{\textbf{Top:}} Host masses of DEBASS, Foundation DR1, and ZTF DR2; \textit{\textbf{Middle:}} global $U-R$ color for DEBASS and Foundation DR1; \textit{\textbf{Bottom:}} directional light radius of DEBASS, Foundation DR1, and ZTF DR2.
\label{fig:hosts}}
\end{figure}

In Figure \ref{fig:hosts}, we examine the host properties of  the overall DEBASS data set, as of mid 2025, as compared to both ZTF and Foundation. Currently, the DEBASS sample features slightly more massive hosts than either of these other programs. The $(U-R)_{\mathrm{global}}$ tends to be distributed more normally than that of Foundation ($U-R$ host color data is not publicly available for ZTF). The distribution of DEBASS's $d_\mathrm{DLR}$ does not differ significantly from Foundation, although it does from ZTF, which features a more significant fraction of data with $d_{DLR} < 1$.

\section{Conclusions}
The anticipated $>400$ cosmologically-useful, low-redshift, DECam-imaged, southern-sky SNe~Ia of DEBASS provide a means of better probing the current state of cosmology, including addressing such questions as whether current experimental suggestions of evolving dark energy is a result of SN systematics. Additionally, DEBASS is a precursor to surveys to surveys like LSST, which will also provide a single-instrument sample of SNe Ia for precision cosmology. Furthermore, with DEBASS observing approved through the first half of 2027, DEBASS has the opportunity to complement and validate on the LSST wide-fast-deep light curves concurrently. 
The 77 DEBASS SNe that constitute this preliminary data release are those that highlight DEBASS's unique ability to dovetail with the DES-SN5YR data, which currently offers one of best sets of cosmological constraints today. 

In this paper, we show the survey structure and analysis methods that are employed across the DEBASS analysis. This work's partner paper \citep{acevedo:inprep-a} goes into further detail on the verification and calibration of DR0.5 and that of the entire DEBASS data set, as well as provides our first cosmology results. Even before bias corrections are applied, our Hubble results using only 62 final SNe~Ia are extremely competitive with both Foundation and ZTF, the first of which consists of SNe that have been used in results suggesting evolving dark energy. 

Moving forward, we will produce both a complete data release consisting of all 400+ DEBASS supernovae and a full cosmological analysis using these, and demonstrate how this sample can be combined with the DES-SN5YR data. 

Future work will supply new constraints on the growth of structure $f\sigma_{8}$, which will benefit from the all-sky uniformity and absolute calibration seen in the DEBASS sample, improving our ability to perform spatially-dependent peculiar velocity field cosmology. We will also place new constraints on the Hubble Constant by examining DEBASS SNe hosted in 8 locally accessible superstructures. Data from these SNe can facilitate an improved calibration of the DESI fundamental plane relation and therefore vastly improve $H_0$ inference.

\section{Acknowledgments}
\begin{acknowledgments}
We thank NOIRLab and DECAT for facilitating remote observations on a 3-day cadence.
We thank Rick Kessler for his advice, support, and guidance in the processing of the DEBASS data.
We thank the Templeton Foundation for directly supporting this research (N. S., D. B., and D. S.). D. S. is supported by Department of Energy grant DE-SC0010007, the David and Lucile Packard Foundation, the Templeton Foundation, and Sloan Foundation. This material is based upon work supported by the National Science Foundation Graduate Research Fellowship under Grant No. DGE 2139754. B. M. is supported by an Australian Government Research Training Program (RTP) Scholarship. We also thank Tamara Davis for the careful reading of this paper.
Based in part on data acquired at the ANU 2.3-metre telescope. The automation of the telescope was made possible through an initial grant provided by the Centre of Gravitational Astrophysics and the Research School of Astronomy and Astrophysics at the Australian National University and through a grant provided by the Australian Research Council through LE230100063. The Lens proposal system is maintained by the AAO Research Data \& Software team as part of the Data Central Science Platform. We acknowledge the traditional custodians of the land on which the telescope stands, the Gamilaraay people, and pay our respects to elders past and present.
\end{acknowledgments}

\vspace{5mm}
\facilities{SSO: WiFeS, CTIO: DECam
}

\bibliography{sample631}{}

\begin{thebibliography}{}
\expandafter\ifx\csname natexlab\endcsname\relax\def\natexlab#1{#1}\fi
\providecommand{\url}[1]{\href{#1}{#1}}
\providecommand{\dodoi}[1]{doi:~\href{http://doi.org/#1}{\nolinkurl{#1}}}
\providecommand{\doeprint}[1]{\href{http://ascl.net/#1}{\nolinkurl{http://ascl.net/#1}}}
\providecommand{\doarXiv}[1]{\href{https://arxiv.org/abs/#1}{\nolinkurl{https://arxiv.org/abs/#1}}}

\bibitem[{{Abdurro'uf} {et~al.}(2022){Abdurro'uf}, {Accetta}, {Aerts}, {Silva Aguirre}, {Ahumada}, {Ajgaonkar}, {Filiz Ak}, {Alam}, {Allende Prieto}, {Almeida}, {Anders}, {Anderson}, {Andrews}, {Anguiano}, {Aquino-Ort{\'\i}z}, {Arag{\'o}n-Salamanca}, {Argudo-Fern{\'a}ndez}, {Ata}, {Aubert}, {Avila-Reese}, {Badenes}, {Barb{\'a}}, {Barger}, {Barrera-Ballesteros}, {Beaton}, {Beers}, {Belfiore}, {Bender}, {Bernardi}, {Bershady}, {Beutler}, {Bidin}, {Bird}, {Bizyaev}, {Blanc}, {Blanton}, {Boardman}, {Bolton}, {Boquien}, {Borissova}, {Bovy}, {Brandt}, {Brown}, {Brownstein}, {Brusa}, {Buchner}, {Bundy}, {Burchett}, {Bureau}, {Burgasser}, {Cabang}, {Campbell}, {Cappellari}, {Carlberg}, {Wanderley}, {Carrera}, {Cash}, {Chen}, {Chen}, {Cherinka}, {Chiappini}, {Choi}, {Chojnowski}, {Chung}, {Clerc}, {Cohen}, {Comerford}, {Comparat}, {da Costa}, {Covey}, {Crane}, {Cruz-Gonzalez}, {Culhane}, {Cunha}, {Dai}, {Damke}, {Darling}, {Davidson}, {Davies}, {Dawson}, {De Lee}, {Diamond-Stanic}, {Cano-D{\'\i}az}, {S{\'a}nchez},
  {Donor}, {Duckworth}, {Dwelly}, {Eisenstein}, {Elsworth}, {Emsellem}, {Eracleous}, {Escoffier}, {Fan}, {Farr}, {Feng}, {Fern{\'a}ndez-Trincado}, {Feuillet}, {Filipp}, {Fillingham}, {Frinchaboy}, {Fromenteau}, {Galbany}, {Garc{\'\i}a}, {Garc{\'\i}a-Hern{\'a}ndez}, {Ge}, {Geisler}, {Gelfand}, {G{\'e}ron}, {Gibson}, {Goddy}, {Godoy-Rivera}, {Grabowski}, {Green}, {Greener}, {Grier}, {Griffith}, {Guo}, {Guy}, {Hadjara}, {Harding}, {Hasselquist}, {Hayes}, {Hearty}, {Hern{\'a}ndez}, {Hill}, {Hogg}, {Holtzman}, {Horta}, {Hsieh}, {Hsu}, {Hsu}, {Huber}, {Huertas-Company}, {Hutchinson}, {Hwang}, {Ibarra-Medel}, {Chitham}, {Ilha}, {Imig}, {Jaekle}, {Jayasinghe}, {Ji}, {Johnson}, {Jones}, {J{\"o}nsson}, {Katkov}, {Khalatyan}, {Kinemuchi}, {Kisku}, {Knapen}, {Kneib}, {Kollmeier}, {Kong}, {Kounkel}, {Kreckel}, {Krishnarao}, {Lacerna}, {Lane}, {Langgin}, {Lavender}, {Law}, {Lazarz}, {Leung}, {Leung}, {Lewis}, {Li}, {Li}, {Lian}, {Liang}, {Lin}, {Lin}, {Lin}, {Lintott}, {Long}, {Longa-Pe{\~n}a}, {L{\'o}pez-Cob{\'a}}, {Lu},
  {Lundgren}, {Luo}, {Mackereth}, {de la Macorra}, {Mahadevan}, {Majewski}, {Manchado}, {Mandeville}, {Maraston}, {Margalef-Bentabol}, {Masseron}, {Masters}, {Mathur}, {McDermid}, {Mckay}, {Merloni}, {Merrifield}, {Meszaros}, {Miglio}, {Di Mille}, {Minniti}, {Minsley}, {Monachesi}, {Moon}, {Mosser}, {Mulchaey}, {Muna}, {Mu{\~n}oz}, {Myers}, {Myers}, {Nadathur}, {Nair}, {Nandra}, {Neumann}, {Newman}, {Nidever}, {Nikakhtar}, {Nitschelm}, {O'Connell}, {Garma-Oehmichen}, {Luan Souza de Oliveira}, {Olney}, {Oravetz}, {Ortigoza-Urdaneta}, {Osorio}, {Otter}, {Pace}, {Padilla}, {Pan}, {Pan}, {Parikh}, {Parker}, {Peirani}, {Pe{\~n}a Ram{\'\i}rez}, {Penny}, {Percival}, {Perez-Fournon}, {Pinsonneault}, {Poidevin}, {Poovelil}, {Price-Whelan}, {B{\'a}rbara de Andrade Queiroz}, {Raddick}, {Ray}, {Rembold}, {Riddle}, {Riffel}, {Riffel}, {Rix}, {Robin}, {Rodr{\'\i}guez-Puebla}, {Roman-Lopes}, {Rom{\'a}n-Z{\'u}{\~n}iga}, {Rose}, {Ross}, {Rossi}, {Rubin}, {Salvato}, {S{\'a}nchez}, {S{\'a}nchez-Gallego}, {Sanderson}, {Santana
  Rojas}, {Sarceno}, {Sarmiento}, {Sayres}, {Sazonova}, {Schaefer}, {Schiavon}, {Schlegel}, {Schneider}, {Schultheis}, {Schwope}, {Serenelli}, {Serna}, {Shao}, {Shapiro}, {Sharma}, {Shen}, {Shetrone}, {Shu}, {Simon}, {Skrutskie}, {Smethurst}, {Smith}, {Sobeck}, {Spoo}, {Sprague}, {Stark}, {Stassun}, {Steinmetz}, {Stello}, {Stone-Martinez}, {Storchi-Bergmann}, {Stringfellow}, {Stutz}, {Su}, {Taghizadeh-Popp}, {Talbot}, {Tayar}, {Telles}, {Teske}, {Thakar}, {Theissen}, {Tkachenko}, {Thomas}, {Tojeiro}, {Hernandez Toledo}, {Troup}, {Trump}, {Trussler}, {Turner}, {Tuttle}, {Unda-Sanzana}, {V{\'a}zquez-Mata}, {Valentini}, {Valenzuela}, {Vargas-Gonz{\'a}lez}, {Vargas-Maga{\~n}a}, {Alfaro}, {Villanova}, {Vincenzo}, {Wake}, {Warfield}, {Washington}, {Weaver}, {Weijmans}, {Weinberg}, {Weiss}, {Westfall}, {Wild}, {Wilde}, {Wilson}, {Wilson}, {Wilson}, {Wolf}, {Wood-Vasey}, {Yan}, {Zamora}, {Zasowski}, {Zhang}, {Zhao}, {Zheng}, {Zheng}, \& {Zhu}}]{SDSS}
{Abdurro'uf}, {Accetta}, K., {Aerts}, C., {et~al.} 2022, \apjs, 259, 35, \dodoi{10.3847/1538-4365/ac4414}

\bibitem[{Acevedo {et~al.}(submitted)Acevedo, Sherman, Scolnic, \& Brout}]{acevedo:inprep-a}
Acevedo, M., Sherman, N.~F., Scolnic, D., \& Brout, D. submitted

\bibitem[{Adame {et~al.}(2025)Adame, Aguilar, Ahlen, Alam, Alexander, Alvarez, Alves, Anand, Andrade, Armengaud, Avila, Aviles, Awan, Bahr-Kalus, Bailey, Baltay, Bault, Behera, BenZvi, Bera, Beutler, Bianchi, Blake, Blum, Brieden, Brodzeller, Brooks, Buckley-Geer, Burtin, Calderon, Canning, Carnero~Rosell, Cereskaite, Cervantes-Cota, Chabanier, Chaussidon, Chaves-Montero, Chen, Chen, Claybaugh, Cole, Cuceu, Davis, Dawson, de~la Macorra, de~Mattia, Deiosso, Dey, Dey, Ding, Doel, Edelstein, Eftekharzadeh, Eisenstein, Elliott, Fagrelius, Fanning, Ferraro, Ereza, Findlay, Flaugher, Font-Ribera, Forero-Sánchez, Forero-Romero, Frenk, Garcia-Quintero, Gaztañaga, Gil-Marín, Gontcho, Gonzalez-Morales, Gonzalez-Perez, Gordon, Green, Gruen, Gsponer, Gutierrez, Guy, Hadzhiyska, Hahn, Hanif, Herrera-Alcantar, Honscheid, Howlett, Huterer, Iršič, Ishak, Juneau, Karaçaylı, Kehoe, Kent, Kirkby, Kremin, Krolewski, Lai, Lan, Landriau, Lang, Lasker, Le~Goff, Le~Guillou, Leauthaud, Levi, Li, Linder, Lodha, Magneville,
  Manera, Margala, Martini, Maus, McDonald, Medina-Varela, Meisner, Mena-Fernández, Miquel, Moon, Moore, Moustakas, Mueller, Muñoz-Gutiérrez, Myers, Nadathur, Napolitano, Neveux, Newman, Nguyen, Nie, Niz, Noriega, Padmanabhan, Paillas, Palanque-Delabrouille, Pan, Penmetsa, Percival, Pieri, Pinon, Poppett, Porredon, Prada, Pérez-Fernández, Pérez-Ràfols, Rabinowitz, Raichoor, Ramírez-Pérez, Ramirez-Solano, Rashkovetskyi, Ravoux, Rezaie, Rich, Rocher, Rockosi, Roe, Rosado-Marin, Ross, Rossi, Ruggeri, Ruhlmann-Kleider, Samushia, Sanchez, Saulder, Schlafly, Schlegel, Schubnell, Seo, Shafieloo, Sharples, Silber, Slosar, Smith, Sprayberry, Tan, Tarlé, Taylor, Trusov, Ureña-López, Vaisakh, Valcin, Valdes, Vargas-Magaña, Verde, Walther, Wang, Wang, Weaver, Weaverdyck, Wechsler, Weinberg, White, Yu, Yu, Yuan, Yèche, Zaborowski, Zarrouk, Zhang, Zhao, Zhao, Zhou, Zhuang, Zou, \& collaboration}]{Adame_2025}
Adame, A., Aguilar, J., Ahlen, S., {et~al.} 2025, Journal of Cosmology and Astroparticle Physics, 2025, 021, \dodoi{10.1088/1475-7516/2025/02/021}

\bibitem[{Albareti {et~al.}(2017)Albareti, Prieto, Almeida, Anders, Anderson, Andrews, Aragón-Salamanca, Argudo-Fernández, Armengaud, Aubourg, Avila-Reese, Badenes, Bailey, Barbuy, Barger, Barrera-Ballesteros, Bartosz, Basu, Bates, Battaglia, Baumgarten, Baur, Bautista, Beers, Belfiore, Bershady, de~Lis, Bird, Bizyaev, Blanc, Blanton, Blomqvist, Bolton, Borissova, Bovy, Brandt, Brinkmann, Brownstein, Bundy, Burtin, Busca, Chavez, Díaz, Cappellari, Carrera, Chen, Cherinka, Cheung, Chiappini, Chojnowski, Chuang, Chung, Cirolini, Clerc, Cohen, Comerford, Comparat, Correa~do Nascimento, Cousinou, Covey, Crane, Croft, Cunha, Darling, Davidson, Dawson, Da~Costa, Da~Silva~Ilha, Machado, Delubac, De~Lee, De~la Macorra, De~la Torre, Diamond-Stanic, Donor, Downes, Drory, Du, Du~Mas~des Bourboux, Dwelly, Ebelke, Eigenbrot, Eisenstein, Elsworth, Emsellem, Eracleous, Escoffier, Evans, Falcón-Barroso, Fan, Favole, Fernandez-Alvar, Fernandez-Trincado, Feuillet, Fleming, Font-Ribera, Freischlad, Frinchaboy, Fu, Gao~高,
  Garcia, Garcia-Dias, Garcia-Hernández, Pérez, Gaulme, Ge, Geisler, Gillespie, Marin, Girardi, Goddard, Chew, Gonzalez-Perez, Grabowski, Green, Grier, Grier, Guo, Guy, Hagen, Hall, Harding, Harley, Hasselquist, Hawley, Hayes, Hearty, Hekker, Toledo, Ho, Hogg, Holley-Bockelmann, Holtzman, Holzer, Hu~胡, Huber, Hutchinson, Hwang, Ibarra-Medel, Ivans, Ivory, Jaehnig, Jensen, Johnson, Jones, Jullo, Kallinger, Kinemuchi, Kirkby, Klaene, Kneib, Kollmeier, Lacerna, Lane, Lang, Laurent, Law, Leauthaud, Le~Goff, Li, Li, Li, Li, Liang~梁, Liang, Lima, Lin~林, Lin~林, Lin~林, Liu, Long, Lucatello, MacDonald, MacLeod, Mackereth, Mahadevan, Maia, Maiolino, Majewski, Malanushenko, Malanushenko, Mallmann, Manchado, Maraston, Marques-Chaves, Valpuesta, Masters, Mathur, McGreer, Merloni, Merrifield, Meszáros, Meza, Miglio, Minchev, Molaverdikhani, Montero-Dorta, Mosser, Muna, Myers, Nair, Nandra, Ness, Newman, Nichol, Nidever, Nitschelm, O’Connell, Oravetz, Oravetz, Pace, Padilla, Palanque-Delabrouille, Pan,
  Parejko, Paris, Park, Peacock, Peirani, Pellejero-Ibanez, Penny, Percival, Percival, Perez-Fournon, Petitjean, Pieri, Pinsonneault, Pisani, Prada, Prakash, Price-Jones, Raddick, Rahman, Raichoor, Rembold, Reyna, Rich, Richstein, Ridl, Riffel, Riffel, Rix, Robin, Rockosi, Rodríguez-Torres, Rodrigues, Roe, Lopes, Román-Zúñiga, Ross, Rossi, Ruan, Ruggeri, Runnoe, Salazar-Albornoz, Salvato, Sanchez, Sanchez, Sanchez-Gallego, Santiago, Schiavon, Schimoia, Schlafly, Schlegel, Schneider, Sch\"{o}nrich, Schultheis, Schwope, Seo, Serenelli, Sesar, Shao, Shetrone, Shull, Aguirre, Skrutskie, Slosar, Smith, Smith, Sobeck, Somers, Souto, Stark, Stassun, Steinmetz, Stello, Bergmann, Strauss, Streblyanska, Stringfellow, Suarez, Sun, Taghizadeh-Popp, Tang, Tao, Tayar, Tembe, Thomas, Tinker, Tojeiro, Tremonti, Troup, Trump, Unda-Sanzana, Valenzuela, Van~den Bosch, Vargas-Magaña, Vazquez, Villanova, Vivek, Vogt, Wake, Walterbos, Wang, Wang, Weaver, Weijmans, Weinberg, Westfall, Whelan, Wilcots, Wild, Williams, Wilson,
  Wood-Vasey, Wylezalek, Xiao~肖, Yan, Yang, Ybarra, Yeche, Yuan, Zakamska, Zamora, Zasowski, Zhang, Zhao, Zhao, Zheng, Zheng, Zhou, Zhu, Zinn, \& Zou}]{Albareti2017}
Albareti, F.~D., Prieto, C.~A., Almeida, A., {et~al.} 2017, The Astrophysical Journal Supplement Series, 233, 25, \dodoi{10.3847/1538-4365/aa8992}

\bibitem[{{Astropy Collaboration} {et~al.}(2013){Astropy Collaboration}, {Robitaille}, {Tollerud}, {Greenfield}, {Droettboom}, {Bray}, {Aldcroft}, {Davis}, {Ginsburg}, {Price-Whelan}, {Kerzendorf}, {Conley}, {Crighton}, {Barbary}, {Muna}, {Ferguson}, {Grollier}, {Parikh}, {Nair}, {Unther}, {Deil}, {Woillez}, {Conseil}, {Kramer}, {Turner}, {Singer}, {Fox}, {Weaver}, {Zabalza}, {Edwards}, {Azalee Bostroem}, {Burke}, {Casey}, {Crawford}, {Dencheva}, {Ely}, {Jenness}, {Labrie}, {Lim}, {Pierfederici}, {Pontzen}, {Ptak}, {Refsdal}, {Servillat}, \& {Streicher}}]{2013A&A...558A..33A}
{Astropy Collaboration}, {Robitaille}, T.~P., {Tollerud}, E.~J., {et~al.} 2013, \aap, 558, A33, \dodoi{10.1051/0004-6361/201322068}

\bibitem[{Becker(2014)}]{githubGitHubAcbeckerhotpants}
Becker, A. 2014, {G}it{H}ub - acbecker/hotpants: hotpants --- github.com, \url{https://github.com/acbecker/hotpants}

\bibitem[{Bellm {et~al.}(2019)Bellm, Kulkarni, Graham, Dekany, Smith, Riddle, Masci, Helou, Prince, Adams, Barbarino, Barlow, Bauer, Beck, Belicki, Biswas, Blagorodnova, Bodewits, Bolin, Brinnel, Brooke, Bue, Bulla, Burruss, Cenko, Chang, Connolly, Coughlin, Cromer, Cunningham, De, Delacroix, Desai, Duev, Eadie, Farnham, Feeney, Feindt, Flynn, Franckowiak, Frederick, Fremling, Gal-Yam, Gezari, Giomi, Goldstein, Golkhou, Goobar, Groom, Hacopians, Hale, Henning, Ho, Hover, Howell, Hung, Huppenkothen, Imel, Ip, Ivezić, Jackson, Jones, Juric, Kasliwal, Kaspi, Kaye, Kelley, Kowalski, Kramer, Kupfer, Landry, Laher, Lee, Lin, Lin, Lunnan, Giomi, Mahabal, Mao, Miller, Monkewitz, Murphy, Ngeow, Nordin, Nugent, Ofek, Patterson, Penprase, Porter, Rauch, Rebbapragada, Reiley, Rigault, Rodriguez, Roestel, Rusholme, Santen, Schulze, Shupe, Singer, Soumagnac, Stein, Surace, Sollerman, Szkody, Taddia, Terek, Van~Sistine, van Velzen, Vestrand, Walters, Ward, Ye, Yu, Yan, \& Zolkower}]{Bellm2018}
Bellm, E.~C., Kulkarni, S.~R., Graham, M.~J., {et~al.} 2019, Publications of the Astronomical Society of the Pacific, 131, 018002, \dodoi{10.1088/1538-3873/aaecbe}

\bibitem[{Bertin(2011)}]{2011ASPC..442..435B}
Bertin, E. 2011, Astronomical Data Analysis Software and Systems XX ASP Conference Series, 442, 435.
\newblock \url{http://aspbooks.org/publications/442/435.pdf}

\bibitem[{Bertin \& Arnouts(1996)}]{Bertin1996}
Bertin, E., \& Arnouts, S. 1996, Astronomy and Astrophysics Supplement Series, 117, 393–404, \dodoi{10.1051/aas:1996164}

\bibitem[{Betoule {et~al.}(2014)Betoule, Kessler, Guy, Mosher, Hardin, Biswas, Astier, El-Hage, Konig, Kuhlmann, Marriner, Pain, Regnault, Balland, Bassett, Brown, Campbell, Carlberg, Cellier-Holzem, Cinabro, Conley, D’Andrea, DePoy, Doi, Ellis, Fabbro, Filippenko, Foley, Frieman, Fouchez, Galbany, Goobar, Gupta, Hill, Hlozek, Hogan, Hook, Howell, Jha, Le~Guillou, Leloudas, Lidman, Marshall, M\"{o}ller, Mourão, Neveu, Nichol, Olmstead, Palanque-Delabrouille, Perlmutter, Prieto, Pritchet, Richmond, Riess, Ruhlmann-Kleider, Sako, Schahmaneche, Schneider, Smith, Sollerman, Sullivan, Walton, \& Wheeler}]{Betoule2014}
Betoule, M., Kessler, R., Guy, J., {et~al.} 2014, Astronomy and Astrophysics, 568, A22, \dodoi{10.1051/0004-6361/201423413}

\bibitem[{Brout \& Scolnic(2021)}]{Brout2021}
Brout, D., \& Scolnic, D. 2021, The Astrophysical Journal, 909, 26, \dodoi{10.3847/1538-4357/abd69b}

\bibitem[{Brout {et~al.}(2019)Brout, Scolnic, Kessler, D’Andrea, Davis, Gupta, Hinton, Kim, Lasker, Lidman, Macaulay, M\"{o}ller, Nichol, Sako, Smith, Sullivan, Zhang, Andersen, Asorey, Avelino, Bassett, Brown, Calcino, Carollo, Challis, Childress, Clocchiatti, Filippenko, Foley, Galbany, Glazebrook, Hoormann, Kasai, Kirshner, Kuehn, Kuhlmann, Lewis, Mandel, March, Miranda, Morganson, Muthukrishna, Nugent, Palmese, Pan, Sharp, Sommer, Swann, Thomas, Tucker, Uddin, Wester, Abbott, Allam, Annis, Avila, Bechtol, Bernstein, Bertin, Brooks, Burke, Rosell, Kind, Carretero, Castander, Cunha, da~Costa, Davis, De~Vicente, DePoy, Desai, Diehl, Doel, Drlica-Wagner, Eifler, Estrada, Fernandez, Flaugher, Fosalba, Frieman, García-Bellido, Gruen, Gruendl, Gutierrez, Hartley, Hollowood, Honscheid, Hoyle, James, Jarvis, Jeltema, Krause, Lahav, Li, Lima, Maia, Marriner, Marshall, Martini, Menanteau, Miller, Miquel, Ogando, Plazas, Romer, Roodman, Rykoff, Sanchez, Santiago, Scarpine, Schubnell, Serrano, Sevilla-Noarbe,
  Smith, Soares-Santos, Sobreira, Suchyta, Swanson, Tarle, Thomas, Troxel, Tucker, Vikram, Walker, \& Zhang}]{Brout2019}
Brout, D., Scolnic, D., Kessler, R., {et~al.} 2019, The Astrophysical Journal, 874, 150, \dodoi{10.3847/1538-4357/ab08a0}

\bibitem[{Brout {et~al.}(2022{\natexlab{a}})Brout, Scolnic, Popovic, Riess, Carr, Zuntz, Kessler, Davis, Hinton, Jones, Kenworthy, Peterson, Said, Taylor, Ali, Armstrong, Charvu, Dwomoh, Meldorf, Palmese, Qu, Rose, Sanchez, Stubbs, Vincenzi, Wood, Brown, Chen, Chambers, Coulter, Dai, Dimitriadis, Filippenko, Foley, Jha, Kelsey, Kirshner, M\"{o}ller, Muir, Nadathur, Pan, Rest, Rojas-Bravo, Sako, Siebert, Smith, Stahl, \& Wiseman}]{Brout2022_a}
Brout, D., Scolnic, D., Popovic, B., {et~al.} 2022{\natexlab{a}}, The Astrophysical Journal, 938, 110, \dodoi{10.3847/1538-4357/ac8e04}

\bibitem[{Brout {et~al.}(2022{\natexlab{b}})Brout, Taylor, Scolnic, Wood, Rose, Vincenzi, Dwomoh, Lidman, Riess, Ali, Qu, \& Dai}]{Brout2022}
Brout, D., Taylor, G., Scolnic, D., {et~al.} 2022{\natexlab{b}}, The Astrophysical Journal, 938, 111, \dodoi{10.3847/1538-4357/ac8bcc}

\bibitem[{Burgaz {et~al.}(2024)Burgaz, Maguire, Dimitriadis, Smith, Sollerman, Galbany, Rigault, Goobar, Johansson, Kim, Alburai, Amenouche, Deckers, Ginolin, Harvey, Muller-Bravo, Nordin, Phan, Rosnet, Nugent, Terwel, Graham, Hale, Kasliwal, Laher, Neill, Purdum, \& Rusholme}]{ztf2}
Burgaz, U., Maguire, K., Dimitriadis, G., {et~al.} 2024, ZTF SN Ia DR2: Properties of the low-mass host galaxies of Type Ia supernovae in a volume-limited sample,  arXiv, \dodoi{10.48550/ARXIV.2412.14262}

\bibitem[{{Campbell} {et~al.}(2016){Campbell}, {Fraser}, \& {Gilmore}}]{Campbell2016}
{Campbell}, H., {Fraser}, M., \& {Gilmore}, G. 2016, \mnras, 457, 3470, \dodoi{10.1093/mnras/stw115}

\bibitem[{{Cappellari} \& {Emsellem}(2004)}]{Cappellari2004}
{Cappellari}, M., \& {Emsellem}, E. 2004, \pasp, 116, 138, \dodoi{10.1086/381875}

\bibitem[{Carr {et~al.}(2022)Carr, Davis, Scolnic, Said, Brout, Peterson, \& Kessler}]{Carr_2022}
Carr, A., Davis, T.~M., Scolnic, D., {et~al.} 2022, Publications of the Astronomical Society of Australia, 39, \dodoi{10.1017/pasa.2022.41}

\bibitem[{Carrick {et~al.}(2015)Carrick, Turnbull, Lavaux, \& Hudson}]{Carrick2015}
Carrick, J., Turnbull, S.~J., Lavaux, G., \& Hudson, M.~J. 2015, Monthly Notices of the Royal Astronomical Society, 450, 317–332, \dodoi{10.1093/mnras/stv547}

\bibitem[{Chambers {et~al.}(2016{\natexlab{a}})Chambers, Magnier, Metcalfe, Flewelling, Huber, Waters, Denneau, Draper, Farrow, Finkbeiner, Holmberg, Koppenhoefer, Price, Rest, Saglia, Schlafly, Smartt, Sweeney, Wainscoat, Burgett, Chastel, Grav, Heasley, Hodapp, Jedicke, Kaiser, Kudritzki, Luppino, Lupton, Monet, Morgan, Onaka, Shiao, Stubbs, Tonry, White, Bañados, Bell, Bender, Bernard, Boegner, Boffi, Botticella, Calamida, Casertano, Chen, Chen, Cole, Deacon, Frenk, Fitzsimmons, Gezari, Gibbs, Goessl, Goggia, Gourgue, Goldman, Grant, Grebel, Hambly, Hasinger, Heavens, Heckman, Henderson, Henning, Holman, Hopp, Ip, Isani, Jackson, Keyes, Koekemoer, Kotak, Le, Liska, Long, Lucey, Liu, Martin, Masci, McLean, Mindel, Misra, Morganson, Murphy, Obaika, Narayan, Nieto-Santisteban, Norberg, Peacock, Pier, Postman, Primak, Rae, Rai, Riess, Riffeser, Rix, R\"{o}ser, Russel, Rutz, Schilbach, Schultz, Scolnic, Strolger, Szalay, Seitz, Small, Smith, Soderblom, Taylor, Thomson, Taylor, Thakar, Thiel, Thilker, Unger,
  Urata, Valenti, Wagner, Walder, Walter, Watters, Werner, Wood-Vasey, \& Wyse}]{panstarrs}
Chambers, K.~C., Magnier, E.~A., Metcalfe, N., {et~al.} 2016{\natexlab{a}}, The Pan-STARRS1 Surveys,  arXiv, \dodoi{10.48550/ARXIV.1612.05560}

\bibitem[{Chambers {et~al.}(2016{\natexlab{b}})Chambers, Magnier, Metcalfe, Flewelling, Huber, Waters, Denneau, Draper, Farrow, Finkbeiner, Holmberg, Koppenhoefer, Price, Rest, Saglia, Schlafly, Smartt, Sweeney, Wainscoat, Burgett, Chastel, Grav, Heasley, Hodapp, Jedicke, Kaiser, Kudritzki, Luppino, Lupton, Monet, Morgan, Onaka, Shiao, Stubbs, Tonry, White, Bañados, Bell, Bender, Bernard, Boegner, Boffi, Botticella, Calamida, Casertano, Chen, Chen, Cole, Deacon, Frenk, Fitzsimmons, Gezari, Gibbs, Goessl, Goggia, Gourgue, Goldman, Grant, Grebel, Hambly, Hasinger, Heavens, Heckman, Henderson, Henning, Holman, Hopp, Ip, Isani, Jackson, Keyes, Koekemoer, Kotak, Le, Liska, Long, Lucey, Liu, Martin, Masci, McLean, Mindel, Misra, Morganson, Murphy, Obaika, Narayan, Nieto-Santisteban, Norberg, Peacock, Pier, Postman, Primak, Rae, Rai, Riess, Riffeser, Rix, R\"{o}ser, Russel, Rutz, Schilbach, Schultz, Scolnic, Strolger, Szalay, Seitz, Small, Smith, Soderblom, Taylor, Thomson, Taylor, Thakar, Thiel, Thilker, Unger,
  Urata, Valenti, Wagner, Walder, Walter, Watters, Werner, Wood-Vasey, \& Wyse}]{pslimmag}
---. 2016{\natexlab{b}}, The Pan-STARRS1 Surveys,  arXiv, \dodoi{10.48550/ARXIV.1612.05560}

\bibitem[{{Childress} {et~al.}(2014){Childress}, {Vogt}, {Nielsen}, \& {Sharp}}]{Childress2013}
{Childress}, M.~J., {Vogt}, F. P.~A., {Nielsen}, J., \& {Sharp}, R.~G. 2014, \apss, 349, 617, \dodoi{10.1007/s10509-013-1682-0}

\bibitem[{Chung {et~al.}(2023)Chung, Yoon, Park, An, Son, Cho, \& Lee}]{Chung2023}
Chung, C., Yoon, S.-J., Park, S., {et~al.} 2023, The Astrophysical Journal, 959, 94, \dodoi{10.3847/1538-4357/ad0121}

\bibitem[{Collaboration: {et~al.}(2016)Collaboration:, Abbott, Abdalla, Aleksić, Allam, Amara, Bacon, Balbinot, Banerji, Bechtol, Benoit-Lévy, Bernstein, Bertin, Blazek, Bonnett, Bridle, Brooks, Brunner, Buckley-Geer, Burke, Caminha, Capozzi, Carlsen, Carnero-Rosell, Carollo, Carrasco-Kind, Carretero, Castander, Clerkin, Collett, Conselice, Crocce, Cunha, D'Andrea, da~Costa, Davis, Desai, Diehl, Dietrich, Dodelson, Doel, Drlica-Wagner, Estrada, Etherington, Evrard, Fabbri, Finley, Flaugher, Foley, Fosalba, Frieman, García-Bellido, Gaztanaga, Gerdes, Giannantonio, Goldstein, Gruen, Gruendl, Guarnieri, Gutierrez, Hartley, Honscheid, Jain, James, Jeltema, Jouvel, Kessler, King, Kirk, Kron, Kuehn, Kuropatkin, Lahav, Li, Lima, Lin, Maia, Makler, Manera, Maraston, Marshall, Martini, McMahon, Melchior, Merson, Miller, Miquel, Mohr, Morice-Atkinson, Naidoo, Neilsen, Nichol, Nord, Ogando, Ostrovski, Palmese, Papadopoulos, Peiris, Peoples, Percival, Plazas, Reed, Refregier, Romer, Roodman, Ross, Rozo, Rykoff, Sadeh,
  Sako, Sánchez, Sanchez, Santiago, Scarpine, Schubnell, Sevilla-Noarbe, Sheldon, Smith, Smith, Soares-Santos, Sobreira, Soumagnac, Suchyta, Sullivan, Swanson, Tarle, Thaler, Thomas, Thomas, Tucker, Vieira, Vikram, Walker, Wechsler, Weller, Wester, Whiteway, Wilcox, Yanny, Zhang, \& Zuntz}]{DES}
Collaboration:, D. E.~S., Abbott, T., Abdalla, F.~B., {et~al.} 2016, Monthly Notices of the Royal Astronomical Society, 460, 1270, \dodoi{10.1093/mnras/stw641}

\bibitem[{Dai {et~al.}(2023)Dai, Jones, Kenworthy, Kessler, Pierel, Foley, Jha, \& Scolnic}]{Dai2023}
Dai, M., Jones, D.~O., Kenworthy, W.~D., {et~al.} 2023, The Astrophysical Journal Supplement Series, 267, 1, \dodoi{10.3847/1538-4365/acd051}

\bibitem[{{DES Collaboration} {et~al.}(2024){DES Collaboration}, Abbott, Acevedo, Aguena, Alarcon, Allam, Alves, Amon, Andrade-Oliveira, Annis, Armstrong, Asorey, Avila, Bacon, Bassett, Bechtol, Bernardinelli, Bernstein, Bertin, Blazek, Bocquet, Brooks, Brout, Buckley-Geer, Burke, Camacho, Camilleri, Campos, Carnero~Rosell, Carollo, Carr, Carretero, Castander, Cawthon, Chang, Chen, Choi, Conselice, Costanzi, da~Costa, Crocce, Davis, DePoy, Desai, Diehl, Dixon, Dodelson, Doel, Doux, Drlica-Wagner, Elvin-Poole, Everett, Ferrero, Ferté, Flaugher, Foley, Fosalba, Friedel, Frieman, Frohmaier, Galbany, García-Bellido, Gatti, Gaztanaga, Giannini, Glazebrook, Graur, Gruen, Gruendl, Gutierrez, Hartley, Herner, Hinton, Hollowood, Honscheid, Huterer, Jain, James, Jeffrey, Kasai, Kelsey, Kent, Kessler, Kim, Kirshner, Kovacs, Kuehn, Lahav, Lee, Lee, Lewis, Li, Lidman, Lin, Malik, Marshall, Martini, Mena-Fernández, Menanteau, Miquel, Mohr, Mould, Muir, M\"{o}ller, Neilsen, Nichol, Nugent, Ogando, Palmese, Pan, Paterno,
  Percival, Pereira, Pieres, Plazas~Malagón, Popovic, Porredon, Prat, Qu, Raveri, Rodríguez-Monroy, Romer, Roodman, Rose, Sako, Sanchez, Sanchez~Cid, Schubnell, Scolnic, Sevilla-Noarbe, Shah, Smith, Smith, Soares-Santos, Suchyta, Sullivan, Suntzeff, Swanson, Sánchez, Tarle, Taylor, Thomas, To, Toy, Troxel, Tucker, Tucker, Uddin, Vincenzi, Walker, Weaverdyck, Wechsler, Weller, Wester, Wiseman, Yamamoto, Yuan, Zhang, \& Zhang}]{Abbott2024}
{DES Collaboration}, Abbott, T. M.~C., Acevedo, M., {et~al.} 2024, The Astrophysical Journal Letters, 973, L14, \dodoi{10.3847/2041-8213/ad6f9f}

\bibitem[{Dhawan {et~al.}(2021)Dhawan, Goobar, Smith, Johansson, Rigault, Nordin, Biswas, Goldstein, Nugent, Kim, Miller, Graham, Medford, Kasliwal, Kulkarni, Duev, Bellm, Rosnet, Riddle, \& Sollerman}]{Dhawan_2021}
Dhawan, S., Goobar, A., Smith, M., {et~al.} 2021, Monthly Notices of the Royal Astronomical Society, 510, 2228–2241, \dodoi{10.1093/mnras/stab3093}

\bibitem[{{Dixon} {et~al.}(2025){Dixon}, {Mould}, {Lidman}, {Taylor}, {Flynn}, {Duffy}, {Galbany}, {Scolnic}, {Davis}, {M{\"o}ller}, {Kelsey}, {Lee}, {Wiseman}, {Vincenzi}, {Shah}, {Aguena}, {Allam}, {Alves}, {Bacon}, {Bocquet}, {Brooks}, {Burke}, {Rosell}, {Carollo}, {Carretero}, {Conselice}, {da Costa}, {Pereira}, {Diehl}, {Doel}, {Everett}, {Ferrero}, {Flaugher}, {Frieman}, {Garc{\'\i}a-Bellido}, {Gatti}, {Gaztanaga}, {Giannini}, {Gruen}, {Gruendl}, {Gutierrez}, {Herner}, {Hinton}, {Hollowood}, {Honscheid}, {James}, {Kuehn}, {Lima}, {Marshall}, {Mena-Fern{\'a}ndez}, {Menanteau}, {Miquel}, {Myles}, {Nichol}, {Ogando}, {Palmese}, {Pieres}, {Malag{\'o}n}, {Samuroff}, {Sanchez}, {Sanchez Cid}, {Sevilla-Noarbe}, {Smith}, {Sobreira}, {Suchyta}, {Swanson}, {Tarle}, {To}, {Tucker}, {Tucker}, {Vikram}, {Walker}, \& {Weaverdyck}}]{Dixon2025}
{Dixon}, M., {Mould}, J., {Lidman}, C., {et~al.} 2025, \mnras, 538, 782, \dodoi{10.1093/mnras/staf266}

\bibitem[{{Dopita} {et~al.}(2007){Dopita}, {Hart}, {McGregor}, {Oates}, {Bloxham}, \& {Jones}}]{Dopita2007}
{Dopita}, M., {Hart}, J., {McGregor}, P., {et~al.} 2007, \apss, 310, 255, \dodoi{10.1007/s10509-007-9510-z}

\bibitem[{Drlica-Wagner {et~al.}(2022)Drlica-Wagner, Ferguson, Adamów, Aguena, Allam, Andrade-Oliveira, Bacon, Bechtol, Bell, Bertin, Bilaji, Bocquet, Bom, Brooks, Burke, Carballo-Bello, Carlin, Carnero~Rosell, Carrasco~Kind, Carretero, Castander, Cerny, Chang, Choi, Conselice, Costanzi, Crnojević, Costa, Vicente, Desai, Esteves, Everett, Ferrero, Fitzpatrick, Flaugher, Friedel, Frieman, García-Bellido, Gatti, Gaztanaga, Gerdes, Gruen, Gruendl, Gschwend, Hartley, Hernandez-Lang, Hinton, Hollowood, Honscheid, Hughes, Jacques, James, Johnson, Kuehn, Kuropatkin, Lahav, Li, Lidman, Lin, March, Marshall, Martínez-Delgado, Martínez-Vázquez, Massana, Mau, McNanna, Melchior, Menanteau, Miller, Miquel, Mohr, Morgan, Mutlu-Pakdil, Muñoz, Neilsen, Nidever, Nikutta, Nilo~Castellon, Noël, Ogando, Olsen, Pace, Palmese, Paz-Chinchón, Pereira, Pieres, Plazas~Malagón, Prat, Riley, Rodriguez-Monroy, Romer, Roodman, Sako, Sakowska, Sanchez, Sánchez, Sand, Santana-Silva, Santiago, Schubnell, Serrano, Sevilla-Noarbe,
  Simon, Smith, Soares-Santos, Stringfellow, Suchyta, Suson, Tan, Tarle, Tavangar, Thomas, To, Tollerud, Troxel, Tucker, Varga, Vivas, Walker, Weller, Wilkinson, Wu, Yanny, Zaborowski, \& Zenteno}]{DW_2022}
Drlica-Wagner, A., Ferguson, P.~S., Adamów, M., {et~al.} 2022, The Astrophysical Journal Supplement Series, 261, 38, \dodoi{10.3847/1538-4365/ac78eb}

\bibitem[{{Fioc} \& {Rocca-Volmerange}(1999)}]{PEGASE}
{Fioc}, M., \& {Rocca-Volmerange}, B. 1999, arXiv e-prints, astro, \dodoi{10.48550/arXiv.astro-ph/9912179}

\bibitem[{{Fitzpatrick}(1999)}]{Fitzpatrick_1999}
{Fitzpatrick}, E.~L. 1999, \pasp, 111, 63, \dodoi{10.1086/316293}

\bibitem[{Flaugher {et~al.}(2015)Flaugher, Diehl, Honscheid, Abbott, Alvarez, Angstadt, Annis, Antonik, Ballester, Beaufore, Bernstein, Bernstein, Bigelow, Bonati, Boprie, Brooks, Buckley-Geer, Campa, Cardiel-Sas, Castander, Castilla, Cease, Cela-Ruiz, Chappa, Chi, Cooper, da~Costa, Dede, Derylo, DePoy, de~Vicente, Doel, Drlica-Wagner, Eiting, Elliott, Emes, Estrada, Fausti~Neto, Finley, Flores, Frieman, Gerdes, Gladders, Gregory, Gutierrez, Hao, Holland, Holm, Huffman, Jackson, James, Jonas, Karcher, Karliner, Kent, Kessler, Kozlovsky, Kron, Kubik, Kuehn, Kuhlmann, Kuk, Lahav, Lathrop, Lee, Levi, Lewis, Li, Mandrichenko, Marshall, Martinez, Merritt, Miquel, Muñoz, Neilsen, Nichol, Nord, Ogando, Olsen, Palaio, Patton, Peoples, Plazas, Rauch, Reil, Rheault, Roe, Rogers, Roodman, Sanchez, Scarpine, Schindler, Schmidt, Schmitt, Schubnell, Schultz, Schurter, Scott, Serrano, Shaw, Smith, Soares-Santos, Stefanik, Stuermer, Suchyta, Sypniewski, Tarle, Thaler, Tighe, Tran, Tucker, Walker, Wang, Watson, Weaverdyck,
  Wester, Woods, \& Yanny}]{Flaugher2015}
Flaugher, B., Diehl, H.~T., Honscheid, K., {et~al.} 2015, The Astronomical Journal, 150, 150, \dodoi{10.1088/0004-6256/150/5/150}

\bibitem[{Foley {et~al.}(2017)Foley, Scolnic, Rest, Jha, Pan, Riess, Challis, Chambers, Coulter, Dettman, Foley, Fox, Huber, Jones, Kilpatrick, Kirshner, Schultz, Siebert, Flewelling, Gibson, Magnier, Miller, Primak, Smartt, Smith, Wainscoat, Waters, \& Willman}]{Foley2017}
Foley, R.~J., Scolnic, D., Rest, A., {et~al.} 2017, Monthly Notices of the Royal Astronomical Society, 475, 193–219, \dodoi{10.1093/mnras/stx3136}

\bibitem[{{Galbany} {et~al.}(2022){Galbany}, {Smith}, {Duarte Puertas}, {Gonz{\'a}lez-Gait{\'a}n}, {Pessa}, {Sako}, {Iglesias-P{\'a}ramo}, {L{\'o}pez-S{\'a}nchez}, {Moll{\'a}}, \& {V{\'\i}lchez}}]{Galbany2022}
{Galbany}, L., {Smith}, M., {Duarte Puertas}, S., {et~al.} 2022, \aap, 659, A89, \dodoi{10.1051/0004-6361/202141568}

\bibitem[{Ginolin {et~al.}(2025)Ginolin, Rigault, Smith, Copin, Ruppin, Dimitriadis, Goobar, Johansson, Maguire, Nordin, Amenouche, Aubert, Barjou-Delayre, Betoule, Burgaz, Carreres, Deckers, Dhawan, Feinstein, Fouchez, Galbany, Ganot, Harvey, de~Jaeger, Kenworthy, Kim, Kowalski, Kuhn, Lacroix, Müller-Bravo, Nugent, Popovic, Racine, Rosnet, Rosselli, Sollerman, Terwel, Townsend, Brugger, Bellm, Kasliwal, Kulkarni, Laher, Masci, Riddle, \& Sharma}]{Ginolin_2025}
Ginolin, M., Rigault, M., Smith, M., {et~al.} 2025, Astronomy and Astrophysics, 695, A140, \dodoi{10.1051/0004-6361/202450378}

\bibitem[{Glazebrook \& Bland‐Hawthorn(2001)}]{Glazebrook_2001}
Glazebrook, K., \& Bland‐Hawthorn, J. 2001, Publications of the Astronomical Society of the Pacific, 113, 197–214, \dodoi{10.1086/318625}

\bibitem[{Goldstein {et~al.}(2015)Goldstein, D’Andrea, Fischer, Foley, Gupta, Kessler, Kim, Nichol, Nugent, Papadopoulos, Sako, Smith, Sullivan, Thomas, Wester, Wolf, Abdalla, Banerji, Benoit-Lévy, Bertin, Brooks, Rosell, Castander, Costa, Covarrubias, DePoy, Desai, Diehl, Doel, Eifler, Neto, Finley, Flaugher, Fosalba, Frieman, Gerdes, Gruen, Gruendl, James, Kuehn, Kuropatkin, Lahav, Li, Maia, Makler, March, Marshall, Martini, Merritt, Miquel, Nord, Ogando, Plazas, Romer, Roodman, Sanchez, Scarpine, Schubnell, Sevilla-Noarbe, Smith, Soares-Santos, Sobreira, Suchyta, Swanson, Tarle, Thaler, \& Walker}]{Goldstein_2015}
Goldstein, D.~A., D’Andrea, C.~B., Fischer, J.~A., {et~al.} 2015, The Astronomical Journal, 150, 82, \dodoi{10.1088/0004-6256/150/3/82}

\bibitem[{Gupta {et~al.}(2016)Gupta, Kuhlmann, Kovacs, Spinka, Kessler, Goldstein, Liotine, Pomian, D’Andrea, Sullivan, Carretero, Castander, Nichol, Finley, Fischer, Foley, Kim, Papadopoulos, Sako, Scolnic, Smith, Tucker, Uddin, Wolf, Yuan, Abbott, Abdalla, Benoit-Lévy, Bertin, Brooks, Rosell, Kind, Cunha, Costa, Desai, Doel, Eifler, Evrard, Flaugher, Fosalba, Gaztañaga, Gruen, Gruendl, James, Kuehn, Kuropatkin, Maia, Marshall, Miquel, Plazas, Romer, Sánchez, Schubnell, Sevilla-Noarbe, Sobreira, Suchyta, Swanson, Tarle, Walker, \& Wester}]{Gupta_2016}
Gupta, R.~R., Kuhlmann, S., Kovacs, E., {et~al.} 2016, The Astronomical Journal, 152, 154, \dodoi{10.3847/0004-6256/152/6/154}

\bibitem[{Herner {et~al.}(2020{\natexlab{a}})Herner, Annis, Brout, Soares-Santos, Kessler, Sako, Butler, Doctor, Palmese, Allam, Tucker, Sobreira, Yanny, Diehl, Frieman, Glaeser, Garcia, Sherman, Bechtol, Berger, Chen, Conselice, Cook, Cowperthwaite, Davis, Drlica-Wagner, Farr, Finley, Foley, Garcia-Bellido, Gill, Gruendl, Holz, Kuropatkin, Lin, Marriner, Marshall, Matheson, Neilsen, Paz-Chinchón, Sauseda, Scolnic, Williams, Avila, Bertin, Buckley-Geer, Burke, Rosell, Carrasco-Kind, Carretero, da~Costa, De~Vicente, Desai, Doel, Eifler, Everett, Fosalba, Gaztanaga, Gerdes, Gschwend, Gutierrez, Hartley, Hollowood, Honscheid, James, Krause, Kuehn, Lahav, Li, Lima, Maia, March, Menanteau, Miquel, Plazas, Sanchez, Scarpine, Schubnell, Serrano, Sevilla-Noarbe, Smith, Suchyta, Tarle, Wester, \& Zhang}]{Herner2020a}
Herner, K., Annis, J., Brout, D., {et~al.} 2020{\natexlab{a}}, Astronomy and Computing, 33, 100425, \dodoi{10.1016/j.ascom.2020.100425}

\bibitem[{Herner {et~al.}(2020{\natexlab{b}})Herner, Annis, Garcia, Soares-Santos, Brout, Glaeser, Sherman, Kessler, Morgan, Palmese, Paz-Chinchon, Lenon, \& Bachmann}]{Herner2020b}
Herner, K., Annis, J., Garcia, A., {et~al.} 2020{\natexlab{b}}, EPJ Web of Conferences, 245, 01008, \dodoi{10.1051/epjconf/202024501008}

\bibitem[{Hinton \& Brout(2020)}]{Hinton2020}
Hinton, S., \& Brout, D. 2020, Journal of Open Source Software, 5, 2122, \dodoi{10.21105/joss.02122}

\bibitem[{Hinton {et~al.}(2016)Hinton, Davis, Lidman, Glazebrook, \& Lewis}]{Hinton2016}
Hinton, S., Davis, T.~M., Lidman, C., Glazebrook, K., \& Lewis, G. 2016, Astronomy and Computing, 15, 61–71, \dodoi{10.1016/j.ascom.2016.03.001}

\bibitem[{Hoaglin {et~al.}(2000)Hoaglin, Mosteller, \& (Editor)}]{Hoaglin00}
Hoaglin, D.~C., Mosteller, F., \& (Editor), J. W.~T. 2000, Understanding Robust and Exploratory Data Analysis, 1st edn. (Wiley-Interscience)

\bibitem[{{Howell} {et~al.}(2005){Howell}, {Sullivan}, {Perrett}, {Bronder}, {Hook}, {Astier}, {Aubourg}, {Balam}, {Basa}, {Carlberg}, {Fabbro}, {Fouchez}, {Guy}, {Lafoux}, {Neill}, {Pain}, {Palanque-Delabrouille}, {Pritchet}, {Regnault}, {Rich}, {Taillet}, {Knop}, {McMahon}, {Perlmutter}, \& {Walton}}]{Howell2005}
{Howell}, D.~A., {Sullivan}, M., {Perrett}, K., {et~al.} 2005, \apj, 634, 1190, \dodoi{10.1086/497119}

\bibitem[{Johansson {et~al.}(2021)Johansson, Cenko, Fox, Dhawan, Stanishev, Butler, Lee, Watson, Fremling, Kasliwal, Nugent, Petrushevska, Sollerman, Yan, Burke, Hosseinzadeh, Howell, McCully, \& Valenti}]{Johansson2021}
Johansson, J., Cenko, S.~B., Fox, O.~D., {et~al.} 2021, The Astrophysical Journal, 923, 237, \dodoi{10.3847/1538-4357/ac2f9e}

\bibitem[{{Jones} {et~al.}(2018){Jones}, {Riess}, {Scolnic}, {Pan}, {Johnson}, {Coulter}, {Dettman}, {Foley}, {Foley}, {Huber}, {Jha}, {Kilpatrick}, {Kirshner}, {Rest}, {Schultz}, \& {Siebert}}]{Jones2018}
{Jones}, D.~O., {Riess}, A.~G., {Scolnic}, D.~M., {et~al.} 2018, \apj, 867, 108, \dodoi{10.3847/1538-4357/aae2b9}

\bibitem[{Jones {et~al.}(2021)Jones, Foley, Narayan, Hjorth, Huber, Aleo, Alexander, Angus, Auchettl, Baldassare, Bruun, Chambers, Chatterjee, Coppejans, Coulter, DeMarchi, Dimitriadis, Drout, Engel, French, Gagliano, Gall, Hung, Izzo, Jacobson-Galán, Kilpatrick, Korhonen, Margutti, Raimundo, Ramirez-Ruiz, Rest, Rojas-Bravo, Siebert, Smartt, Smith, Terreran, Wang, Wojtak, Agnello, Ansari, Arendse, Baldeschi, Blanchard, Brethauer, Bright, Brown, Boer, Dodd, Fairlamb, Grillo, Hajela, Cold, Kolborg, Law-Smith, Lin, Magnier, Malanchev, Matthews, Mockler, Muthukrishna, Pan, Pfister, Ramanah, Rest, Sarangi, Schrøder, Stauffer, Stroh, Taggart, Tinyanont, \& Wainscoat}]{Jones2021}
Jones, D.~O., Foley, R.~J., Narayan, G., {et~al.} 2021, The Astrophysical Journal, 908, 143, \dodoi{10.3847/1538-4357/abd7f5}

\bibitem[{Kelsey {et~al.}(2020)Kelsey, Sullivan, Smith, Wiseman, Brout, Davis, Frohmaier, Galbany, Grayling, Gutiérrez, Hinton, Kessler, Lidman, M\"{o}ller, Sako, Scolnic, Uddin, Vincenzi, Abbott, Aguena, Allam, Annis, Avila, Bacon, Bertin, Brooks, Burke, Carnero Rosell, Carrasco Kind, Carretero, Castander, Costanzi, da Costa, Desai, Diehl, Doel, Everett, Ferrero, Ferté, Flaugher, Fosalba, García-Bellido, Gerdes, Gruen, Gruendl, Gschwend, Gutierrez, Hollowood, Honscheid, James, Kim, Kuehn, Kuropatkin, Lahav, Lima, Marshall, Martini, Menanteau, Miquel, Morgan, Ogando, Palmese, Paz-Chinchón, Plazas, Romer, Sánchez, Sanchez, Serrano, Sevilla-Noarbe, Suchyta, Tarle, Thomas, To, Varga, Walker, \& Wilkinson}]{Kelsey2020}
Kelsey, L., Sullivan, M., Smith, M., {et~al.} 2020, Monthly Notices of the Royal Astronomical Society, 501, 4861–4876, \dodoi{10.1093/mnras/staa3924}

\bibitem[{{Kelsey} {et~al.}(2021){Kelsey}, {Sullivan}, {Smith}, {Wiseman}, {Brout}, {Davis}, {Frohmaier}, {Galbany}, {Grayling}, {Guti{\'e}rrez}, {Hinton}, {Kessler}, {Lidman}, {M{\"o}ller}, {Sako}, {Scolnic}, {Uddin}, {Vincenzi}, {Abbott}, {Aguena}, {Allam}, {Annis}, {Avila}, {Bacon}, {Bertin}, {Brooks}, {Burke}, {Carnero Rosell}, {Carrasco Kind}, {Carretero}, {Castander}, {Costanzi}, {da Costa}, {Desai}, {Diehl}, {Doel}, {Everett}, {Ferrero}, {Fert{\'e}}, {Flaugher}, {Fosalba}, {Garc{\'\i}a-Bellido}, {Gerdes}, {Gruen}, {Gruendl}, {Gschwend}, {Gutierrez}, {Hollowood}, {Honscheid}, {James}, {Kim}, {Kuehn}, {Kuropatkin}, {Lahav}, {Lima}, {Marshall}, {Martini}, {Menanteau}, {Miquel}, {Morgan}, {Ogando}, {Palmese}, {Paz-Chinch{\'o}n}, {Plazas}, {Romer}, {S{\'a}nchez}, {Sanchez}, {Serrano}, {Sevilla-Noarbe}, {Suchyta}, {Tarle}, {Thomas}, {To}, {Varga}, {Walker}, {Wilkinson}, \& {DES Collaboration}}]{Kelsey2021}
{Kelsey}, L., {Sullivan}, M., {Smith}, M., {et~al.} 2021, \mnras, 501, 4861, \dodoi{10.1093/mnras/staa3924}

\bibitem[{Kelsey {et~al.}(2023)Kelsey, Sullivan, Wiseman, Armstrong, Chen, Brout, Davis, Dixon, Frohmaier, Galbany, Graur, Kessler, Lidman, M\"{o}ller, Popovic, Rose, Scolnic, Smith, Vincenzi, Abbott, Aguena, Allam, Alves, Annis, Bacon, Bertin, Bocquet, Brooks, Burke, Carnero Rosell, CarrascoKind, Carretero, Costanzi, da Costa, Pereira, Desai, Diehl, Everett, Ferrero, Frieman, García-Bellido, Gruen, Gruendl, Gschwend, Gutierrez, Hinton, Hollowood, Honscheid, James, Kuehn, Kuropatkin, Lewis, Mena-Fernández, Miquel, Palmese, Paz-Chinchón, Pieres, Plazas Malagón, Raveri, Rodriguez-Monroy, Romer, Sanchez, Scarpine, Schubnell, Sevilla-Noarbe, Suchyta, Swanson, Tarle, Tucker, \& Weaverdyck}]{Kelsey2023}
Kelsey, L., Sullivan, M., Wiseman, P., {et~al.} 2023, Monthly Notices of the Royal Astronomical Society, 519, 3046–3063, \dodoi{10.1093/mnras/stac3711}

\bibitem[{{Kenworthy} {et~al.}(2021){Kenworthy}, {Jones}, {Dai}, {Kessler}, {Scolnic}, {Brout}, {Siebert}, {Pierel}, {Dettman}, {Dimitriadis}, {Foley}, {Jha}, {Pan}, {Riess}, {Rodney}, \& {Rojas-Bravo}}]{Kenworthy21}
{Kenworthy}, W.~D., {Jones}, D.~O., {Dai}, M., {et~al.} 2021, \apj, 923, 265, \dodoi{10.3847/1538-4357/ac30d8}

\bibitem[{Kessler {et~al.}(2009)Kessler, Bernstein, Cinabro, Dilday, Frieman, Jha, Kuhlmann, Miknaitis, Sako, Taylor, \& Vanderplas}]{Kessler2009}
Kessler, R., Bernstein, J.~P., Cinabro, D., {et~al.} 2009, Publications of the Astronomical Society of the Pacific, 121, 1028–1035, \dodoi{10.1086/605984}

\bibitem[{Kessler {et~al.}(2015)Kessler, Marriner, Childress, Covarrubias, D’Andrea, Finley, Fischer, Foley, Goldstein, Gupta, Kuehn, Marcha, Nichol, Papadopoulos, Sako, Scolnic, Smith, Sullivan, Wester, Yuan, Abbott, Abdalla, Allam, Benoit-Lévy, Bernstein, Bertin, Brooks, Rosell, Kind, Castander, Crocce, Costa, Desai, Diehl, Eifler, Neto, Flaugher, Frieman, Gerdes, Gruen, Gruendl, Honscheid, James, Kuropatkin, Li, Maia, Marshall, Martini, Miller, Miquel, Nord, Ogando, Plazas, Reil, Romer, Roodman, Sanchez, Sevilla-Noarbe, Smith, Soares-Santos, Sobreira, Tarle, Thaler, Thomas, Tucker, \& Walker}]{Kessler2015}
Kessler, R., Marriner, J., Childress, M., {et~al.} 2015, The Astronomical Journal, 150, 172, \dodoi{10.1088/0004-6256/150/6/172}

\bibitem[{Knox \& Millea(2020)}]{Knox2020}
Knox, L., \& Millea, M. 2020, Physical Review D, 101, \dodoi{10.1103/physrevd.101.043533}

\bibitem[{Kron(1980)}]{Kron1980}
Kron, R.~G. 1980, The Astrophysical Journal Supplement Series, 43, 305, \dodoi{10.1086/190669}

\bibitem[{{Kroupa}(2001)}]{Kroupa_2001}
{Kroupa}, P. 2001, \mnras, 322, 231, \dodoi{10.1046/j.1365-8711.2001.04022.x}

\bibitem[{Lavaux \& Hudson(2011)}]{Lavaux2011}
Lavaux, G., \& Hudson, M.~J. 2011, Monthly Notices of the Royal Astronomical Society, 416, 2840–2856, \dodoi{10.1111/j.1365-2966.2011.19233.x}

\bibitem[{{Lidman} {et~al.}(2020){Lidman}, {Tucker}, {Davis}, {Uddin}, {Asorey}, {Bolejko}, {Brout}, {Calcino}, {Carollo}, {Carr}, {Childress}, {Hoormann}, {Foley}, {Galbany}, {Glazebrook}, {Hinton}, {Kessler}, {Kim}, {King}, {Kremin}, {Kuehn}, {Lagattuta}, {Lewis}, {Macaulay}, {Malik}, {March}, {Martini}, {M{\"o}ller}, {Mudd}, {Nichol}, {Panther}, {Parkinson}, {Pursiainen}, {Sako}, {Swann}, {Scalzo}, {Scolnic}, {Sharp}, {Smith}, {Sommer}, {Sullivan}, {Webb}, {Wiseman}, {Yu}, {Yuan}, {Zhang}, {Abbott}, {Aguena}, {Allam}, {Annis}, {Avila}, {Bertin}, {Bhargava}, {Brooks}, {Carnero Rosell}, {Carrasco Kind}, {Carretero}, {Castander}, {Costanzi}, {da Costa}, {De Vicente}, {Doel}, {Eifler}, {Everett}, {Fosalba}, {Frieman}, {Garc{\'\i}a-Bellido}, {Gaztanaga}, {Gruen}, {Gruendl}, {Gschwend}, {Gutierrez}, {Hartley}, {Hollowood}, {Honscheid}, {James}, {Kuropatkin}, {Li}, {Lima}, {Lin}, {Maia}, {Marshall}, {Melchior}, {Menanteau}, {Miquel}, {Palmese}, {Paz-Chinch{\'o}n}, {Plazas}, {Roodman}, {Rykoff}, {Sanchez},
  {Santiago}, {Scarpine}, {Schubnell}, {Serrano}, {Sevilla-Noarbe}, {Suchyta}, {Swanson}, {Tarle}, {Tucker}, {Varga}, {Walker}, {Wester}, {Wilkinson}, \& {DES Collaboration}}]{Lidman2020}
{Lidman}, C., {Tucker}, B.~E., {Davis}, T.~M., {et~al.} 2020, \mnras, 496, 19, \dodoi{10.1093/mnras/staa1341}

\bibitem[{Martin {et~al.}(2024)Martin, Lidman, Brout, Tucker, Dixon, \& Armstrong}]{Martin2024}
Martin, B., Lidman, C., Brout, D., {et~al.} 2024, Monthly Notices of the Royal Astronomical Society, 533, 2640–2655, \dodoi{10.1093/mnras/stae1996}

\bibitem[{Morgan {et~al.}(2020)Morgan, Soares-Santos, Annis, Herner, Garcia, Palmese, Drlica-Wagner, Kessler, García-Bellido, Bachmann, Sherman, Allam, Bechtol, Bom, Brout, Butler, Butner, Cartier, Chen, Conselice, Cook, Davis, Doctor, Farr, Figueiredo, Finley, Foley, Galarza, Gill, Gruendl, Holz, Kuropatkin, Lidman, Lin, Malik, Mann, Marriner, Marshall, Martínez-Vázquez, Meza, Neilsen, Nicolaou, Olivares~E., Paz-Chinchón, Points, Quirola-Vásquez, Rodriguez, Sako, Scolnic, Smith, Sobreira, Tucker, Vivas, Wiesner, Wood, Yanny, Zenteno, Abbott, Aguena, Avila, Bertin, Bhargava, Brooks, Burke, Rosell, Kind, Carretero, Costa, Costanzi, De~Vicente, Desai, Diehl, Doel, Eifler, Everett, Flaugher, Frieman, Gaztanaga, Gerdes, Gruen, Gschwend, Gutierrez, Hartley, Hinton, Hollowood, Honscheid, James, Kuehn, Lahav, Lima, Maia, March, Miquel, Ogando, Plazas, Roodman, Sanchez, Scarpine, Schubnell, Serrano, Sevilla-Noarbe, Suchyta, \& Tarle}]{Morgan2020}
Morgan, R., Soares-Santos, M., Annis, J., {et~al.} 2020, The Astrophysical Journal, 901, 83, \dodoi{10.3847/1538-4357/abafaa}

\bibitem[{Morganson {et~al.}(2018)Morganson, Gruendl, Menanteau, Kind, Chen, Daues, Drlica-Wagner, Friedel, Gower, Johnson, Johnson, Kessler, Paz-Chinchón, Petravick, Pond, Yanny, Allam, Armstrong, Barkhouse, Bechtol, Benoit-Lévy, Bernstein, Bertin, Buckley-Geer, Covarrubias, Desai, Diehl, Goldstein, Gruen, Li, Lin, Marriner, Mohr, Neilsen, Ngeow, Paech, Rykoff, Sako, Sevilla-Noarbe, Sheldon, Sobreira, Tucker, \& Wester}]{Morganson2018}
Morganson, E., Gruendl, R.~A., Menanteau, F., {et~al.} 2018, Publications of the Astronomical Society of the Pacific, 130, 074501, \dodoi{10.1088/1538-3873/aab4ef}

\bibitem[{Müller-Bravo \& Galbany(2022)}]{Hostphot}
Müller-Bravo, T.~E., \& Galbany, L. 2022, J. open source softw., 7, 4508, \dodoi{10.21105/joss.04508}

\bibitem[{{Pan} {et~al.}(2014){Pan}, {Sullivan}, {Maguire}, {Hook}, {Nugent}, {Howell}, {Arcavi}, {Botyanszki}, {Cenko}, {DeRose}, {Fakhouri}, {Gal-Yam}, {Hsiao}, {Kulkarni}, {Laher}, {Lidman}, {Nordin}, {Walker}, \& {Xu}}]{Pan2014}
{Pan}, Y.~C., {Sullivan}, M., {Maguire}, K., {et~al.} 2014, \mnras, 438, 1391, \dodoi{10.1093/mnras/stt2287}

\bibitem[{Peterson {et~al.}(2022)Peterson, Kenworthy, Scolnic, Riess, Brout, Carr, Courtois, Davis, Dwomoh, Jones, Popovic, Rose, \& Said}]{Peterson2022}
Peterson, E.~R., Kenworthy, W.~D., Scolnic, D., {et~al.} 2022, The Astrophysical Journal, 938, 112, \dodoi{10.3847/1538-4357/ac4698}

\bibitem[{Pierel {et~al.}(2022)Pierel, Jones, Kenworthy, Dai, Kessler, Ashall, Do, Peterson, Shappee, Siebert, Barna, Brink, Burke, Calamida, Camacho-Neves, Jaeger, Filippenko, Foley, Galbany, Fox, Gomez, Hiramatsu, Hounsell, Howell, Jha, Kwok, Pérez-Fournon, Poidevin, Rest, Rubin, Scolnic, Shirley, Strolger, Tinyanont, \& Wang}]{Pierel2022}
Pierel, J. D.~R., Jones, D.~O., Kenworthy, W.~D., {et~al.} 2022, The Astrophysical Journal, 939, 11, \dodoi{10.3847/1538-4357/ac93f9}

\bibitem[{{Planck Collaboration} {et~al.}(2020){Planck Collaboration}, Aghanim, Akrami, Ashdown, Aumont, Baccigalupi, Ballardini, Banday, Barreiro, Bartolo, Basak, Battye, Benabed, Bernard, Bersanelli, Bielewicz, Bock, Bond, Borrill, Bouchet, Boulanger, Bucher, Burigana, Butler, Calabrese, Cardoso, Carron, Challinor, Chiang, Chluba, Colombo, Combet, Contreras, Crill, Cuttaia, de~Bernardis, de~Zotti, Delabrouille, Delouis, Di~Valentino, Diego, Doré, Douspis, Ducout, Dupac, Dusini, Efstathiou, Elsner, Enßlin, Eriksen, Fantaye, Farhang, Fergusson, Fernandez-Cobos, Finelli, Forastieri, Frailis, Fraisse, Franceschi, Frolov, Galeotta, Galli, Ganga, Génova-Santos, Gerbino, Ghosh, González-Nuevo, Górski, Gratton, Gruppuso, Gudmundsson, Hamann, Handley, Hansen, Herranz, Hildebrandt, Hivon, Huang, Jaffe, Jones, Karakci, Keih\"{a}nen, Keskitalo, Kiiveri, Kim, Kisner, Knox, Krachmalnicoff, Kunz, Kurki-Suonio, Lagache, Lamarre, Lasenby, Lattanzi, Lawrence, Le~Jeune, Lemos, Lesgourgues, Levrier, Lewis, Liguori, Lilje,
  Lilley, Lindholm, López-Caniego, Lubin, Ma, Macías-Pérez, Maggio, Maino, Mandolesi, Mangilli, Marcos-Caballero, Maris, Martin, Martinelli, Martínez-González, Matarrese, Mauri, McEwen, Meinhold, Melchiorri, Mennella, Migliaccio, Millea, Mitra, Miville-Desch\^enes, Molinari, Montier, Morgante, Moss, Natoli, Nørgaard-Nielsen, Pagano, Paoletti, Partridge, Patanchon, Peiris, Perrotta, Pettorino, Piacentini, Polastri, Polenta, Puget, Rachen, Reinecke, Remazeilles, Renzi, Rocha, Rosset, Roudier, Rubiño-Martín, Ruiz-Granados, Salvati, Sandri, Savelainen, Scott, Shellard, Sirignano, Sirri, Spencer, Sunyaev, Suur-Uski, Tauber, Tavagnacco, Tenti, Toffolatti, Tomasi, Trombetti, Valenziano, Valiviita, Van~Tent, Vibert, Vielva, Villa, Vittorio, Wandelt, Wehus, White, White, Zacchei, \& Zonca}]{planck2020}
{Planck Collaboration}, Aghanim, N., Akrami, Y., {et~al.} 2020, Astronomy and Astrophysics, 641, A6, \dodoi{10.1051/0004-6361/201833910}

\bibitem[{Popovic {et~al.}(2022)Popovic, Brout, Kessler, \& Scolnic}]{popovic2022pantheonanalysisforwardmodelingdust}
Popovic, B., Brout, D., Kessler, R., \& Scolnic, D. 2022, The Pantheon+ Analysis: Forward-Modeling the Dust and Intrinsic Colour Distributions of Type Ia Supernovae, and Quantifying their Impact on Cosmological Inferences.
\newblock \doarXiv{2112.04456}

\bibitem[{Popovic {et~al.}(2024)Popovic, Wiseman, Sullivan, Smith, González-Gaitán, Scolnic, Duarte, Armstrong, Asorey, Brout, Carollo, Galbany, Glazebrook, Kelsey, Kessler, Lidman, Lee, Lewis, Möller, Nichol, Sánchez, Toy, Tucker, Vincenzi, Abbott, Aguena, Andrade-Oliveira, Bacon, Brooks, Burke, Rosell, Carretero, Castander, da~Costa, Pereira, Davis, Desai, Everett, Ferrero, Flaugher, García-Bellido, Gaztanaga, Gruendl, Gutierrez, Hinton, Hollowood, Honscheid, James, Kuehn, Lahav, Lee, Marshall, Mena-Fernández, Miquel, Myles, Ogando, Palmese, Pieres, Malagón, Sanchez, Cid, Schubnell, Sevilla-Noarbe, Suchyta, Swanson, Tarle, Vikram, \& Weaverdyck}]{popovic2024modellingimpacthostgalaxy}
Popovic, B., Wiseman, P., Sullivan, M., {et~al.} 2024, Modelling the impact of host galaxy dust on type Ia supernova distance measurements.
\newblock \doarXiv{2406.05051}

\bibitem[{Qu {et~al.}(2024)Qu, Sako, Vincenzi, Sánchez, Brout, Kessler, Chen, Davis, Galbany, Kelsey, Lee, Lidman, Popovic, Rose, Scolnic, Smith, Sullivan, Wiseman, Abbott, Aguena, Alves, Bacon, Bertin, Brooks, Burke, Carnero~Rosell, Carretero, da~Costa, Pereira, Diehl, Doel, Everett, Ferrero, Frieman, García-Bellido, Giannini, Gruen, Gruendl, Gutierrez, Hinton, Hollowood, Honscheid, James, Kuehn, Lahav, Marshall, Mena-Fernández, Menanteau, Miquel, Ogando, Palmese, Pieres, Plazas-Malagón, Raveri, Sanchez, Sevilla-Noarbe, Soares-Santos, Suchyta, Tarle, \& Weaverdyck}]{Qu2024}
Qu, H., Sako, M., Vincenzi, M., {et~al.} 2024, The Astrophysical Journal, 964, 134, \dodoi{10.3847/1538-4357/ad251d}

\bibitem[{Riess {et~al.}(2022)Riess, Yuan, Macri, Scolnic, Brout, Casertano, Jones, Murakami, Anand, Breuval, Brink, Filippenko, Hoffmann, Jha, D’arcy~Kenworthy, Mackenty, Stahl, \& Zheng}]{Riess2022}
Riess, A.~G., Yuan, W., Macri, L.~M., {et~al.} 2022, The Astrophysical Journal Letters, 934, L7, \dodoi{10.3847/2041-8213/ac5c5b}

\bibitem[{{Rigault} {et~al.}(2020){Rigault}, {Brinnel}, {Aldering}, {Antilogus}, {Aragon}, {Bailey}, {Baltay}, {Barbary}, {Bongard}, {Boone}, {Buton}, {Childress}, {Chotard}, {Copin}, {Dixon}, {Fagrelius}, {Feindt}, {Fouchez}, {Gangler}, {Hayden}, {Hillebrandt}, {Howell}, {Kim}, {Kowalski}, {Kuesters}, {Leget}, {Lombardo}, {Lin}, {Nordin}, {Pain}, {Pecontal}, {Pereira}, {Perlmutter}, {Rabinowitz}, {Runge}, {Rubin}, {Saunders}, {Smadja}, {Sofiatti}, {Suzuki}, {Taubenberger}, {Tao}, \& {Thomas}}]{Rigault2020}
{Rigault}, M., {Brinnel}, V., {Aldering}, G., {et~al.} 2020, \aap, 644, A176, \dodoi{10.1051/0004-6361/201730404}

\bibitem[{Rigault {et~al.}(2024{\natexlab{a}})Rigault, Smith, Goobar, Maguire, Dimitriadis, Burgaz, Dhawan, Sollerman, Regnault, Kowalski, Amenouche, Aubert, Barjou-Delayre, Bautista, Bloom, Carreres, Chen, Copin, Deckers, Fouchez, Fremling, Galbany, Ginolin, Graham, Kasliwal, Kenworthy, Kim, Kuhn, Masci, M\"{u}ller-Bravo, Miller, Johansson, Nordin, Nugent, Andreoni, Bellm, Betoule, Osman, Perley, Popovic, Rosnet, Rosselli, Ruppin, Senzel, Rusholme, Schweyer, Terwel, Townsend, Tzanidakis, Wold, Purdum, Qin, Racine, Reusch, Riddle, \& Yan}]{ztfov}
Rigault, M., Smith, M., Goobar, A., {et~al.} 2024{\natexlab{a}}, ZTF SN Ia DR2: Overview,  arXiv, \dodoi{10.48550/ARXIV.2409.04346}

\bibitem[{Rigault {et~al.}(2024{\natexlab{b}})Rigault, Smith, Goobar, Maguire, Dimitriadis, Burgaz, Dhawan, Sollerman, Regnault, Kowalski, Amenouche, Aubert, Barjou-Delayre, Bautista, Bloom, Carreres, Chen, Copin, Deckers, Fouchez, Fremling, Galbany, Ginolin, Graham, Kasliwal, Kenworthy, Kim, Kuhn, Masci, M\"{u}ller-Bravo, Miller, Johansson, Nordin, Nugent, Andreoni, Bellm, Betoule, Osman, Perley, Popovic, Rosnet, Rosselli, Ruppin, Senzel, Rusholme, Schweyer, Terwel, Townsend, Tzanidakis, Wold, Purdum, Qin, Racine, Reusch, Riddle, \& Yan}]{ztfother}
---. 2024{\natexlab{b}}, ZTF SN Ia DR2: Overview,  arXiv, \dodoi{10.48550/ARXIV.2409.04346}

\bibitem[{{Roman} {et~al.}(2018){Roman}, {Hardin}, {Betoule}, {Astier}, {Balland}, {Ellis}, {Fabbro}, {Guy}, {Hook}, {Howell}, {Lidman}, {Mitra}, {M{\"o}ller}, {Mour{\~a}o}, {Neveu}, {Palanque-Delabrouille}, {Pritchet}, {Regnault}, {Ruhlmann-Kleider}, {Saunders}, \& {Sullivan}}]{Roman2018}
{Roman}, M., {Hardin}, D., {Betoule}, M., {et~al.} 2018, \aap, 615, A68, \dodoi{10.1051/0004-6361/201731425}

\bibitem[{{Rose} {et~al.}(2021){Rose}, {Rubin}, {Strolger}, \& {Garnavich}}]{Rose2021}
{Rose}, B.~M., {Rubin}, D., {Strolger}, L., \& {Garnavich}, P.~M. 2021, \apj, 909, 28, \dodoi{10.3847/1538-4357/abd550}

\bibitem[{Rubin {et~al.}(2023)Rubin, Aldering, Betoule, Fruchter, Huang, Kim, Lidman, Linder, Perlmutter, Ruiz-Lapuente, \& Suzuki}]{union3}
Rubin, D., Aldering, G., Betoule, M., {et~al.} 2023, Union Through UNITY: Cosmology with 2, 000 SNe Using a Unified Bayesian Framework,  arXiv, \dodoi{10.48550/ARXIV.2311.12098}

\bibitem[{Rykoff {et~al.}(2023)Rykoff, Tucker, Burke, Allam, Bechtol, Bernstein, Brout, Gruendl, Lasker, Smith, Wester, Yanny, Abbott, Aguena, Alves, Andrade-Oliveira, Annis, Bacon, Bertin, Brooks, Rosell, Carretero, Castander, Choi, da~Costa, Pereira, Davis, De~Vicente, Diehl, Doel, Drlica-Wagner, Everett, Ferrero, Frieman, García-Bellido, Giannini, Gruen, Gutierrez, Hinton, Hollowood, James, Kuehn, Lahav, Marshall, Mena-Fernández, Menanteau, Myles, Nord, Ogando, Palmese, Pieres, Malagón, Raveri, Rodgríguez-Monroy, Sanchez, Santiago, Schubnell, Sevilla-Noarbe, Smith, Soares-Santos, Suchyta, Swanson, Varga, Vincenzi, Walker, Weaverdyck, \& Wiseman}]{rykoffdesy6}
Rykoff, E.~S., Tucker, D.~L., Burke, D.~L., {et~al.} 2023, The Dark Energy Survey Six-Year Calibration Star Catalog,  arXiv, \dodoi{10.48550/ARXIV.2305.01695}

\bibitem[{Sako {et~al.}(2018)Sako, Bassett, Becker, Brown, Campbell, Wolf, Cinabro, D’Andrea, Dawson, DeJongh, Depoy, Dilday, Doi, Filippenko, Fischer, Foley, Frieman, Galbany, Garnavich, Goobar, Gupta, Hill, Hayden, Hlozek, Holtzman, Hopp, Jha, Kessler, Kollatschny, Leloudas, Marriner, Marshall, Miquel, Morokuma, Mosher, Nichol, Nordin, Olmstead, \"{O}stman, Prieto, Richmond, Romani, Sollerman, Stritzinger, Schneider, Smith, Wheeler, Yasuda, \& Zheng}]{Sako2018}
Sako, M., Bassett, B., Becker, A.~C., {et~al.} 2018, Publications of the Astronomical Society of the Pacific, 130, 064002, \dodoi{10.1088/1538-3873/aab4e0}

\bibitem[{{Schlafly} \& {Finkbeiner}(2011)}]{Schlafly_Finkbeiner}
{Schlafly}, E.~F., \& {Finkbeiner}, D.~P. 2011, \apj, 737, 103, \dodoi{10.1088/0004-637X/737/2/103}

\bibitem[{Scolnic {et~al.}(2022)Scolnic, Brout, Carr, Riess, Davis, Dwomoh, Jones, Ali, Charvu, Chen, Peterson, Popovic, Rose, Wood, Brown, Chambers, Coulter, Dettman, Dimitriadis, Filippenko, Foley, Jha, Kilpatrick, Kirshner, Pan, Rest, Rojas-Bravo, Siebert, Stahl, \& Zheng}]{Scolnic2022}
Scolnic, D., Brout, D., Carr, A., {et~al.} 2022, The Astrophysical Journal, 938, 113, \dodoi{10.3847/1538-4357/ac8b7a}

\bibitem[{Scolnic {et~al.}(2018)Scolnic, Jones, Rest, Pan, Chornock, Foley, Huber, Kessler, Narayan, Riess, Rodney, Berger, Brout, Challis, Drout, Finkbeiner, Lunnan, Kirshner, Sanders, Schlafly, Smartt, Stubbs, Tonry, Wood-Vasey, Foley, Hand, Johnson, Burgett, Chambers, Draper, Hodapp, Kaiser, Kudritzki, Magnier, Metcalfe, Bresolin, Gall, Kotak, McCrum, \& Smith}]{Scolnic_2018}
Scolnic, D.~M., Jones, D.~O., Rest, A., {et~al.} 2018, The Astrophysical Journal, 859, 101, \dodoi{10.3847/1538-4357/aab9bb}

\bibitem[{Sherman(2024)}]{thesis}
Sherman, N. 2024, PhD thesis, \dodoi{10.7302/24003}

\bibitem[{Smith {et~al.}(2020)Smith, Smartt, Young, Tonry, Denneau, Flewelling, Heinze, Weiland, Stalder, Rest, Stubbs, Anderson, Chen, Clark, Do, F\"{o}rster, Fulton, Gillanders, McBrien, O’Neill, Srivastav, \& Wright}]{Smith2020}
Smith, K.~W., Smartt, S.~J., Young, D.~R., {et~al.} 2020, Publications of the Astronomical Society of the Pacific, 132, 085002, \dodoi{10.1088/1538-3873/ab936e}

\bibitem[{{Steinhardt} {et~al.}(2020){Steinhardt}, {Sneppen}, \& {Sen}}]{Steinhardt2020}
{Steinhardt}, C.~L., {Sneppen}, A., \& {Sen}, B. 2020, \apj, 902, 14, \dodoi{10.3847/1538-4357/abb140}

\bibitem[{{Sullivan} {et~al.}(2006){Sullivan}, {Le Borgne}, {Pritchet}, {Hodsman}, {Neill}, {Howell}, {Carlberg}, {Astier}, {Aubourg}, {Balam}, {Basa}, {Conley}, {Fabbro}, {Fouchez}, {Guy}, {Hook}, {Pain}, {Palanque-Delabrouille}, {Perrett}, {Regnault}, {Rich}, {Taillet}, {Baumont}, {Bronder}, {Ellis}, {Filiol}, {Lusset}, {Perlmutter}, {Ripoche}, \& {Tao}}]{Sullivan_2006}
{Sullivan}, M., {Le Borgne}, D., {Pritchet}, C.~J., {et~al.} 2006, \apj, 648, 868, \dodoi{10.1086/506137}

\bibitem[{Sullivan {et~al.}(2010)Sullivan, Conley, Howell, Neill, Astier, Balland, Basa, Carlberg, Fouchez, Guy, Hardin, Hook, Pain, Palanque-Delabrouille, Perrett, Pritchet, Regnault, Rich, Ruhlmann-Kleider, Baumont, Hsiao, Kronborg, Lidman, Perlmutter, \& Walker}]{Sullivan2010}
Sullivan, M., Conley, A., Howell, D.~A., {et~al.} 2010, Monthly Notices of the Royal Astronomical Society, no, \dodoi{10.1111/j.1365-2966.2010.16731.x}

\bibitem[{Sánchez {et~al.}(2024)Sánchez, Brout, Vincenzi, Sako, Herner, Kessler, Davis, Scolnic, Acevedo, Lee, Möller, Qu, Kelsey, Wiseman, Armstrong, Rose, Camilleri, Chen, Galbany, Kovacs, Lidman, Popovic, Smith, Sullivan, Toy, Carollo, Glazebrook, Lewis, Nichol, Tucker, Abbott, Aguena, Allam, Alves, Annis, Asorey, Avila, Bacon, Brooks, Burke, Rosell, Carretero, Castander, da~Costa, Duarte, Pereira, Desai, Diehl, Everett, Ferrero, Flaugher, Frieman, García-Bellido, Gatti, Gaztanaga, Giannini, Gruendl, Gutierrez, Hinton, Hollowood, Honscheid, James, Kuehn, Lahav, Lee, Lin, Marshall, Mena-Fernández, Miquel, Myles, Palmese, Pieres, Malagón, Porredon, Romer, Sanchez, Cid, Sevilla-Noarbe, Suchyta, Swanson, Tarle, Tucker, Vikram, Walker, \& Weaverdyck}]{sanchez2024darkenergysurveysupernova}
Sánchez, B.~O., Brout, D., Vincenzi, M., {et~al.} 2024, The Dark Energy Survey Supernova Program: Light curves and 5-Year data release.
\newblock \doarXiv{2406.05046}

\bibitem[{Taylor {et~al.}(2023)Taylor, Jones, Popovic, Vincenzi, Kessler, Scolnic, Dai, Kenworthy, \& Pierel}]{Taylor_2023}
Taylor, G., Jones, D.~O., Popovic, B., {et~al.} 2023, Monthly Notices of the Royal Astronomical Society, 520, 5209–5224, \dodoi{10.1093/mnras/stad320}

\bibitem[{Tonry {et~al.}(2018)Tonry, Denneau, Heinze, Stalder, Smith, Smartt, Stubbs, Weiland, \& Rest}]{Tonry2018}
Tonry, J.~L., Denneau, L., Heinze, A.~N., {et~al.} 2018, Publications of the Astronomical Society of the Pacific, 130, 064505, \dodoi{10.1088/1538-3873/aabadf}

\bibitem[{{Tripp}(1998)}]{1998A&A...331..815T}
{Tripp}, R. 1998, \aap, 331, 815

\bibitem[{{Vazdekis} {et~al.}(2016){Vazdekis}, {Koleva}, {Ricciardelli}, {R{\"o}ck}, \& {Falc{\'o}n-Barroso}}]{Vazdekis2016}
{Vazdekis}, A., {Koleva}, M., {Ricciardelli}, E., {R{\"o}ck}, B., \& {Falc{\'o}n-Barroso}, J. 2016, \mnras, 463, 3409, \dodoi{10.1093/mnras/stw2231}

\bibitem[{Vincenzi {et~al.}(2024{\natexlab{a}})Vincenzi, Brout, Armstrong, Popovic, Taylor, Acevedo, Camilleri, Chen, Davis, Hinton, Kelsey, Kessler, Lee, Lidman, M\"{o}ller, Qu, Sako, Sanchez, Scolnic, Smith, Sullivan, Wiseman, Asorey, Bassett, Carollo, Carr, Foley, Frohmaier, Galbany, Glazebrook, Graur, Kovacs, Kuehn, Malik, Nichol, Rose, Tucker, Toy, Tucker, Yuan, Abbott, Aguena, Alves, Andrade-Oliveira, Annis, Bacon, Bechtol, Bernstein, Brooks, Burke, Rosell, Carretero, Castander, Conselice, da~Costa, Pereira, Desai, Diehl, Doel, Ferrero, Flaugher, Friedel, Frieman, García-Bellido, Gatti, Giannini, Gruen, Gruendl, Hollowood, Honscheid, Huterer, James, Kuropatkin, Lahav, Lee, Lin, Marshall, Mena-Fernández, Menanteau, Miquel, Palmese, Pieres, Malagón, Porredon, Romer, Roodman, Sanchez, Cid, Schubnell, Sevilla-Noarbe, Suchyta, Swanson, Tarle, To, Walker, \& Weaverdyck}]{mariav2024}
Vincenzi, M., Brout, D., Armstrong, P., {et~al.} 2024{\natexlab{a}}, The Dark Energy Survey Supernova Program: Cosmological Analysis and Systematic Uncertainties,  arXiv, \dodoi{10.48550/ARXIV.2401.02945}

\bibitem[{Vincenzi {et~al.}(2024{\natexlab{b}})Vincenzi, Brout, Armstrong, Popovic, Taylor, Acevedo, Camilleri, Chen, Davis, Lee, Lidman, Hinton, Kelsey, Kessler, M\"{o}ller, Qu, Sako, Sanchez, Scolnic, Smith, Sullivan, Wiseman, Asorey, Bassett, Carollo, Carr, Foley, Frohmaier, Galbany, Glazebrook, Graur, Kovacs, Kuehn, Malik, Nichol, Rose, Tucker, Toy, Tucker, Yuan, Abbott, Aguena, Alves, Allam, Andrade-Oliveira, Annis, Bacon, Bechtol, Bernstein, Brooks, Burke, Carnero~Rosell, Carretero, Castander, Conselice, da~Costa, Pereira, Desai, Diehl, Doel, Ferrero, Flaugher, Friedel, Frieman, García-Bellido, Gatti, Giannini, Gruen, Gruendl, Hollowood, Honscheid, Huterer, James, Kuropatkin, Lahav, Lee, Lin, Marshall, Mena-Fernández, Menanteau, Miquel, Palmese, Pieres, Plazas~Malagón, Porredon, Romer, Roodman, Sanchez, Sanchez~Cid, Schubnell, Sevilla-Noarbe, Suchyta, Swanson, Tarle, To, Walker, Weaverdyck, \& Yamamoto}]{Vincenzi2024}
---. 2024{\natexlab{b}}, The Astrophysical Journal, 975, 86, \dodoi{10.3847/1538-4357/ad5e6c}

\bibitem[{Vincenzi {et~al.}(2025)Vincenzi, Kessler, Shah, Lee, Davis, Scolnic, Armstrong, Brout, Camilleri, Chen, Galbany, Lidman, Möller, Popovic, Rose, Sako, Sánchez, Smith, Sullivan, Wiseman, Abbott, Aguena, Allam, Andrade-Oliveira, Bocquet, Brooks, Rosell, Carretero, da~Costa, Pereira, Diehl, Doel, Everett, Flaugher, Frieman, García-Bellido, Gaztanaga, Gruen, Gruendl, Gutierrez, Hinton, Hollowood, Honscheid, James, Kuehn, Lahav, Lee, Marshall, Mena-Fernández, Miquel, Muir, Myles, Palmese, Malagón, Porredon, Samuroff, Sanchez, Cid, Sevilla-Noarbe, Suchyta, Tarle, To, Tucker, Vikram, Walker, Weaverdyck, Weller, \& Collaboration}]{vincenzi2025comparingdessn5yrpantheonsn}
Vincenzi, M., Kessler, R., Shah, P., {et~al.} 2025, Comparing the DES-SN5YR and Pantheon+ SN cosmology analyses: Investigation based on "Evolving Dark Energy or Supernovae systematics?".
\newblock \doarXiv{2501.06664}

\bibitem[{Wiseman {et~al.}(2023)Wiseman, Sullivan, Smith, \& Popovic}]{Wiseman2023}
Wiseman, P., Sullivan, M., Smith, M., \& Popovic, B. 2023, Monthly Notices of the Royal Astronomical Society, 520, 6214–6222, \dodoi{10.1093/mnras/stad488}

\bibitem[{Wiseman {et~al.}(2020)Wiseman, Smith, Childress, Kelsey, M\"{o}ller, Gupta, Swann, Angus, Brout, Davis, Foley, Frohmaier, Galbany, Gutiérrez, Inserra, Kessler, Lewis, Lidman, Macaulay, Nichol, Pursiainen, Sako, Scolnic, Sommer, Sullivan, Tucker, Abbott, Aguena, Allam, Avila, Bertin, Brooks, Buckley-Geer, Burke, Carnero Rosell, Carollo, Carrasco Kind, da Costa, De Vicente, Desai, Diehl, Doel, Eifler, Everett, Fosalba, Frieman, García-Bellido, Gaztanaga, Gerdes, Gill, Glazebrook, Gruendl, Gschwend, Hartley, Hinton, Hollowood, Honscheid, James, Kuehn, Kuropatkin, Lima, Maia, March, Martini, Melchior, Menanteau, Miquel, Ogando, Paz-Chinchón, Plazas, Romer, Roodman, Sanchez, Scarpine, Serrano, Suchyta, Swanson, Tarle, Thomas, Tucker, Varga, Walker, \& Wilkinson}]{Wiseman2020}
Wiseman, P., Smith, M., Childress, M., {et~al.} 2020, Monthly Notices of the Royal Astronomical Society, 495, 4040–4060, \dodoi{10.1093/mnras/staa1302}

\end{thebibliography}
\bibliographystyle{aasjournal}

\begin{longtable*}{  p{1.25cm} p{1.5cm} p{1.55cm} p{1.25cm} p{1.25cm} p{1.25cm} p{1cm} p{1.25cm} p{1.25cm} p{2cm} }
\hline
SN & RA (deg) & DEC (deg) & $z_{\mathrm{CMB}}$ & $z_{\mathrm{err}}$ & $\log_{10}{(M_{\mathrm{Host}}/M_{\odot})}$ & $v_\mathrm{pec}$ & $\mu$ & $\mu_{\mathrm{err}}$ & Discovery Group \\
\hline
2021aabw & 82.40942 & -23.99264 & 0.03237 & 1e-05 & 10.02 & 206 & 16.5116 & 0.05304 & ATLAS \\
2021aafz & 31.07767 & -44.75298 & 0.06625 & 1e-05 & 9.826 & 31 & 17.98224 & 0.05731 & ATLAS \\
2021aaga & 84.95372 & -40.97721 & 0.03812 & 1e-05 & 9.426 & 155 &   &   & ATLAS \\
2021abmp & 11.17652 & -38.18385 & 0.02229 & 1e-05 & 9.216 & -41 & 15.70844 & 0.14669 & ATLAS \\
2021aclv & 14.48071 & -5.13177 & 0.01729 & 1e-05 & 10.565 & -23 &   &   & ZTF \\
2021acxb & 49.9028 & -27.49403 & 0.06372 & 1e-05 & 10.992 & 160 &   &   & Pan-STARRS \\
2021aefx & 64.97217 & -54.94794 & 0.00501 & 1e-05 & 10.411 & -117 &   &   & DLT40\footnote{\citealp{Tartaglia2018}} \\
2021aexy & 48.97436 & -23.04482 & 0.04426 & 1e-05 & 8.616 & -132 & 17.2432 & 0.05446 & ZTF \\
2021agbv & 29.16772 & -14.13008 & 0.05097 & 1e-05 & 9.324 & -158 & 17.7041 & 0.04638 & ATLAS \\
2021clw & 87.83308 & -50.58254 & 0.02369 & 0.00015 & 10.503 & 233 & 15.68216 & 0.05807 & ASAS-SN\footnote{\citealp{Shappee2014}} \\
2021gmk & 91.86167 & -45.00414 & 0.04171 & 0.00015 & 10.557 & 125 & 16.74825 & 0.05196 & ATLAS \\
2021jad & 83.34238 & -21.95186 & 0.0061 & 1e-05 & 10.383 & -97 & 13.27543 & 0.12145 & ATLAS \\
2021ymn & 77.77311 & -17.92368 & 0.05069 & 1e-05 & 10.414 & -37 & 17.32431 & 0.06386 & ZTF \\
2021ywa & 57.85254 & -17.32874 & 0.0672 & 1e-05 & 10.953 & -11 & 18.14871 & 0.28606 & ZTF \\
2021zdx & 46.76907 & -15.42628 & 0.07136 & 1e-05 & 10.366 & -26 & 18.24439 & 0.06287 & Pan-STARRS \\
2021zfr & 45.13148 & -14.95571 & 0.0314 & 1e-05 & 9.045 & -207 & 16.38179 & 0.05521 & ZTF \\
2021zgf & 26.34688 & -42.70028 & 0.05528 & 1e-05 & 10.125 & 112 & 17.53793 & 0.07115 & ATLAS \\
2021zqo & 44.77528 & -18.67142 & 0.06128 & 1e-05 & 10.574 & 41 & 17.81813 & 0.08198 & ZTF \\
2022abgg & 35.75371 & -9.19697 & 0.06951 & 1e-05 & 8.225 & -53 & 18.44991 & 0.09352 & ATLAS \\
2022abic & 44.7887 & -4.47118 & 0.01397 & 0.00015 & 10.37 & -10 & 14.9082 & 0.05974 & ATLAS \\
2022abid & 31.83117 & -25.4407 & 0.01418 & 1e-05 & 10.263 & -19 & 14.64716 & 0.06235 & ATLAS \\
2022aeed & 61.89157 & -31.47193 & 0.072 & 1e-05 & 10.746 & 34 & 18.21102 & 0.06152 & ATLAS \\
2022qke & 16.3515 & -2.31252 & 0.07461 & 1e-05 & 10.584 & -76 & 18.44613 & 0.26308 & ZTF \\
2022qsn & 40.74935 & -19.1946 & 0.05973 & 1e-05 & 11.207 & -12 & 17.98475 & 0.07412 & ZTF \\
2022qwx & 30.5334 & -20.4061 & 0.03123 & 1e-05 & 9.153 & -57 & 16.4143 & 0.15759 & ZTF \\
2022rnu & 25.18103 & -14.47878 & 0.05559 & 1e-05 & 9.28 & -285 & 17.7319 & 0.0592 & ZTF \\
2022ssk & 37.36896 & -5.84837 & 0.04705 & 0.00015 & 9.96 & -149 & 17.32093 & 0.07876 & ZTF \\
2022szg & 33.04323 & -6.21086 & 0.05883 & 0.00015 & 9.082 & -264 & 17.80069 & 0.18143 & ZTF \\
2022wxo & 29.07362 & -7.69947 & 0.0551 & 1e-05 & 10.423 & -280 & 17.98068 & 0.08849 & ZTF \\
2022xhd & 54.88416 & -17.66439 & 0.04205 & 1e-05 & 10.165 & -301 & 17.04096 & 0.06327 & ZTF \\
2022yvv & 40.42924 & -22.81426 & 0.02462 & 1e-05 & 9.579 & -114 & 15.93515 & 0.06538 &  ZTF \\
2022ywf & 20.55082 & 0.95658 & 0.00691 & 0.00015 & 11.382 & -137 &   &   &  ZTF \\
2023aajf & 65.67305 & -51.48758 & 0.04275 & 1e-05 & 11.169 & 151 & 16.9635 & 0.06155 &  GOTO\footnote{\citealp{10.1117/12.2311865}} \\
2023abjs & 54.71622 & -34.36328 & 0.06367 & 1e-05 & 9.742 & -9 &   &   &  ATLAS \\
2023ctf & 65.43506 & -38.72333 & 0.04076 & 1e-05 & 9.392 & 138 & 16.91689 & 0.06559 &  ATLAS \\
2023cul & 88.44887 & -34.59782 & 0.03948 & 1e-05 & 9.685 & 65 & 16.77395 & 0.0682 &  ATLAS \\
2023E & 18.98232 & -50.18747 & 0.02402 & 1e-05 & 10.582 & -117 & 15.50341 & 0.10299 &  ATLAS \\
2023fpa & 87.70845 & -33.09378 & 0.0259 & 1e-05 & 8.674 & 278 & 15.84479 & 0.05063 & ATLAS \\
2023mha & 33.83819 & -24.6036 & 0.03681 & 1e-05 & 10.659 & -146 & 16.72852 & 0.10949 & ATLAS \\
2023mkg & 30.4297 & -21.77317 & 0.0437 & 1e-05 & 9.899 & -20 &   &   & ATLAS \\
2023mmb & 48.07997 & -24.61433 & 0.04039 & 1e-05 & 11.03 & -127 & 16.89203 & 0.07431 & ATLAS \\
2023ndf & 22.81225 & -17.70383 & 0.01684 & 0.00015 & 9.296 & -76 & 15.08275 & 0.05746 & ATLAS \\
2023ngb & 6.4237 & -1.05031 & 0.06993 & 0.00015 & 10.614 & -83 &   &   & ZTF \\
2023oho & 37.51946 & -10.4512 & 0.0116 & 0.00015 & 9.284 & 59 & 14.07737 & 0.06467 & ATLAS \\
2023ono & 43.60764 & -22.50268 & 0.02814 & 1e-05 & 10.374 & -50 &   &   & ATLAS \\
2023ons & 331.28009 & -45.41464 & 0.0629 & 1e-05 & 10.971 & -111 & 17.87863 & 0.12869 & ATLAS \\
2023ptg & 39.14186 & -5.01163 & 0.02917 & 1e-05 & 10.287 & -122 & 16.35005 & 0.06167 & ZTF \\
2023pwl & 327.02531 & -50.56361 & 0.01656 & 1e-05 & 10.6 & -47 &   &   & ATLAS \\
2023qov & 318.00848 & -49.25502 & 0.00937 & 0.00015 & 9.669 & 62 & 13.75072 & 0.11198 & ATLAS \\
2023snj & 22.05926 & -56.49344 & 0.0515 & 1e-05 & 10.139 & 109 &   &   & ATLAS \\
2023tfg & 333.54438 & -52.72263 & 0.07399 & 1e-05 & 10.657 & 79 &   &   & ATLAS \\
2023tfj & 73.68131 & -46.14647 & 0.05777 & 1e-05 & 10.325 & -112 & 17.54488 & 0.08674 & ATLAS \\
2023tnb & 18.21541 & -23.82145 & 0.06106 & 1e-05 & 10.726 & -147 & 17.82781 & 0.06658 & ZTF \\
2023vcr & 68.14435 & -32.52232 & 0.06252 & 1e-05 & 10.87 & -147 & 17.99176 & 0.11602 & ATLAS \\
2023yzu & 80.99785 & -20.09254 & 0.04444 & 1e-05 & 10.705 & -131 & 17.05409 & 0.05352 & Pan-STARRS \\
2023yzw & 19.56777 & -1.00405 & 0.04619 & 0.00015 & 9.807 & -202 & 17.15832 & 0.07748 & Pan-STARRS \\
2023zbo & 25.60522 & -16.29855 & 0.01834 & 0.00015 & 10.238 & -216 & 15.00484 & 0.0604 & Pan-STARRS \\
2024abtt & 46.47638 & -12.27455 & 0.02989 & 0.00015 & 9.299 & -26 & 16.16082 & 0.05327 & ZTF \\
2024acac & 84.12958 & -42.30826 & 0.067 & 0.00015 & 9.362 & 89 & 17.96408 & 0.05531 & GOTO \\
2024acde & 29.01217 & -2.44313 & 0.05535 & 0.00015 & 9.783 & -184 & 17.69068 & 0.06261 & ATLAS \\
2024adzq & 18.87778 & -32.68753 & 0.05662 & 0.00015 & 9.722 & 76 & 17.62603 & 0.06159 & GOTO \\
2024adzv & 37.57407 & -29.38328 & 0.05967 & 0.00015 & 10.865 & 160 &   &   & GOTO \\
2024aemk & 87.26015 & -20.05974 & 0.05766 & 0.00015 & 9.648 & -140 & 17.72674 & 0.05955 & ATLAS \\
2024aemo & 64.41549 & -30.05995 & 0.05795 & 0.00015 & 10.314 & 181 & 17.61364 & 0.07605 & GOTO \\
2024agf & 31.83855 & 1.92845 & 0.02458 & 1e-05 & 10.682 & -182 & 15.65213 & 0.41436 & ZTF \\
2024bha & 82.8916 & -39.05147 & 0.0777 & 1e-05 & 11.073 & 72 & 18.38638 & 0.06427 & GOTO \\
2024hl & 88.68446 & -35.32041 & 0.03538 & 1e-05 & 10.651 & 274 & 16.53964 & 0.05964 & ATLAS \\
2024uns & 38.70142 & -7.68361 & 0.02136 & 1e-05 & 10.369 & -90 &   &   &  ZTF \\
2024vjo & 27.45843 & -36.343 & 0.03275 & 1e-05 & 9.79 & -108 & 16.58834 & 0.05384 &  GOTO \\
2024vke & 49.34801 & -15.17654 & 0.03019 & 1e-05 & 7.813 & -36 & 16.22877 & 0.07344 &  ATLAS \\
2024vsf & 72.06897 & -20.59436 & 0.07365 & 1e-05 & 10.613 & 15 & 18.05253 & 0.07937 &  ZTF \\
2024wuj & 39.58494 & 4.72318 & 0.07431 & 1e-05 & 9.819 & -85 & 18.39156 & 0.07666 &  ATLAS \\
2024xal & 56.00565 & -14.36233 & 0.00504 & 1e-05 & 9.557 & -101 &   &   &  Pan-STARRS \\
2024xyn & 39.84987 & 2.65219 & 0.02064 & 1e-05 & 9.515 & 68 & 15.70081 & 0.05307 & ATLAS \\
2024yid & 41.82074 & -69.2039 & 0.06023 & 1e-05 & 11.29136 & 179 & 17.61323 & 0.05952 & ATLAS \\
2024ykr & 17.63535 & -5.70538 & 0.03621 & 1e-05 & 10.283 & 40 & 16.59045 & 0.06006 & ATLAS \\
2024zsj & 52.18215 & -19.95229 & 0.04537 & 1e-05 & 10.747 & -106 & 17.19344 & 0.0646 & GOTO \\
\hline

\caption{Details of the 77 DEBASS SNe falling within the DES footprint and the subset of 62 cosmology SNe.\label{snedets}}
\end{longtable*}

\end{document}